\documentclass[iop]{emulateapj-rtx4}
\usepackage{hyperref}
\usepackage{bm}
\usepackage{threeparttable}
\usepackage{CJK}
\usepackage{calligra}
\usepackage[T1]{fontenc}

\def\andname{ }

\def\bm{}

\def\bfnabla{\ensuremath{\nabla}}
\def\kms{\,\mbox{km s}^{-1}}

\def\msun{M_\odot}

\def\rsun{R_\odot}

\renewcommand\bv{\boldsymbol{ v}}
\newcommand\bb{\boldsymbol{ B}}
\newcommand\bF{\boldsymbol{ F}}
\newcommand\bP{\boldsymbol{ P}}

\newcommand{\kb}{\mathrm{k_B}}
\newcommand{\cF}{{\scriptsize \calligra F}}
\newcommand{\code}[1]{\texttt{#1}}
\newcommand{\mesa}{\code{MESA}}

\def\<{\,\langle\langle}
\def\>{\,\rangle\rangle}

\shorttitle{Radiation MHD Simulations of Massive Star Envelopes}

\begin{document}
\begin{CJK*}{UTF8}{gbsn}

\shortauthors{Y.-F. Jiang et al.}
\author{Yan-Fei Jiang (姜燕飞)\altaffilmark{1}, Matteo Cantiello\altaffilmark{1}, Lars Bildsten\altaffilmark{1,2}, Eliot Quataert\altaffilmark{3}, 
Omer Blaes\altaffilmark{2}}
\affil{$^1$Kavli Institute for Theoretical Physics, University of California, Santa Barbara, CA 93106, USA} 
\affil{$^2$Department of Physics, University of California, Santa Barbara, CA 93106, USA}
\affil{$^3$Astronomy Department and Theoretical Astrophysics Center, 
University of California at Berkeley, Berkeley, CA 94720-3411, USA}

\title{The Effects of Magnetic Fields on the Dynamics of Radiation Pressure Dominated 
Massive Star Envelopes}

\begin{abstract}
We use three dimensional radiation magneto-hydrodynamic simulations
to study the effects of magnetic fields on the energy transport and
structure of radiation pressure dominated main sequence 
massive star envelopes at the 
region of the iron opacity peak. We focus on the regime 
where the local thermal timescale is shorter than
the dynamical timescale, corresponding to inefficient convective
energy transport. 
We begin with initially weak magnetic fields
relative to the thermal pressure, from $100-1000 {\rm G}$ in differing
geometries. The unstable density inversion amplifies the magnetic field, increasing
the magnetic energy density to values close to equipartition with the
turbulent kinetic energy density. By providing pressure support, the
magnetic field's presence significantly increases the density
fluctuations in the turbulent envelope, thereby enhancing the
radiative energy transport by allowing photons to diffuse out through
low density regions.  Magnetic buoyancy brings small scale magnetic fields to the 
photosphere and increases the vertical
energy transport with the energy advection
velocity proportional to the Alfv\'en velocity, although in all cases
we study photon diffusion still dominates the energy transport. The
increased radiative and advective energy transport causes the stellar
envelope to shrink by several scale heights. We also 
find larger turbulent velocity fluctuations compared to the purely
hydrodynamic case, reaching $\approx 100\kms$ at the stellar
photosphere. The photosphere also shows vertical oscillations 
with similar averaged velocities and periods of a few hours. 
The increased turbulent velocity 
and oscillations will have strong impacts on the line broadening and 
periodic signals in massive stars. 





\end{abstract}

\maketitle

\section{Introduction}

Magnetic fields in the range $100\rm G$ to $20\rm kG$ have been observed at the surface 
of 5-10\% of O-type stars based on high-resolution, sensitive spectropolarimeters \citep[][]{DonatiLandstreet2009,Wadeetal2014,Wadeetal2016}. 
These fields are mostly dipolar and do not show rapid evolution, or any 
systematic correlations with stellar properties such as mass or rotation \citep[][]{LandstreetMathys2000,DonatiLandstreet2009,Landstreetetal2008,Wadeetal2016}.
One limitation of Zeeman spectro-polarimetry applied to unresolved stars, is that it is only sensitive to the mean component of the magnetic field, making the 
technique essentially blind to small scale fields below a certain amplitude \citep[][]{Schnerr:2008,Kochukhov:2013}. There exists indirect evidence that magnetic fields at the surface of massive stars could be much more common than revealed by current spectro-polarimetric observations, and could explain ubiquitous phenomena such as line profile variability \citep{Fullerton:1996} and discrete absorption components in UV spectra \citep{Kaper:1994,Cranmer:1996}.

Dynamo action in the rotating radiative regions of these stars has been suggested on theoretical grounds \citep{Spruit2002,Mullan:2005}, and equipartition magnetic 
fields could be generated by dynamo action in the sub-surface convective regions of massive stars  \citep{Cantiello2009,CantielloBraithwaite2011,Cantiello:2011}.
During the main sequence, the most prominent sub-surface convective regions are located around the iron opacity peak, 
at temperatures $\approx 1.8\times 10^5$ K \citep[][]{Maeder:2008,Cantiello2009,Paxtonetal2013}, where 
the opacity is enhanced by a factor of few compared with the electron scattering value 
and causes the local radiation acceleration to be larger than the gravitational acceleration. 
In this situation, analytical calculations and one dimensional stellar evolution models  predict the development of a density inversion \citep[][]{Joss:1973,Paxtonetal2013}. 

\cite{Jiangetal2015} carried out the first three dimensional radiation hydrodynamic simulations around this region, showing that 
this density inversion is convectively unstable. 
When the iron opacity peak is close to the surface, 
where the optical depth per pressure scale height $\tau_0$ is smaller than $\tau_c$, 
the ratio  between the speed of light and the isothermal sound speed,  convection is inefficient.
The density inversion is not completely erased and, even when the envelope experiences  strong density fluctuations 
and vertical oscillations, it still persists in a time averaged sense. 
In this paper we study how these results are 
affected by the presence of magnetic fields. 


Magnetic fields are also interesting because they provide the opportunity for other instabilities to develop in the stellar envelope, even in the absence of rotation. One 
example is the photon bubble instability \citep[][]{Arons1992,Gammie1998}, an overstability where 
the motion of the gas along the magnetic field lines is amplified by the radiation flux, leading to growing, propagating 
density variations \citep[][]{BlaesSocrates2001}. Analytic work and numerical experiments \citep[][]{Begelman2001,Turneretal2005,Jiangetal2012} 
show that the photon bubble instability can result in shock trains propagating through the envelope.

Convection is also expected to be affected   by the presence of a magnetic
field, as the magnetic pressure for a 1 kG magnetic field
is already comparable to the thermal pressure at the iron opacity peak.
The standard convective instability associated with the iron opacity peak can turn into
the interchange mode of the Parker instability, and the undulatory Parker
instability also becomes possible  if the magnetic energy density decreases
outward \citep[][]{Newcomb1961,Gilman1970,Acheson1979}.

The goal of this paper is to study 
the structure and the energy transport in radiation pressure dominated massive star envelopes, 
based on self-consistent three-dimensional (3D) radiation 
magneto-hydrodynamic simulations. 
In Section \ref{sec:method}, we describe the numerical method we use, which is a direct extension of the 
radiation hydrodynamic algorithm used in \cite{Jiangetal2015}  including  magnetic fields. The simulation 
setup and parameters we choose are also given in Section \ref{sec:method}. The main results are summarized 
in Sections \ref{sec:result} and \ref{sec:result2}. Implications of the simulation results for stellar evolution models and observations 
are given in Section \ref{sec:discussion}.
 
%

\section{Numerical Method}
\label{sec:method}

\subsection{Equations}
Following \cite{Jiangetal2015}, we model plane parallel  
massive star envelopes in cartesian coordinates $(x,y,z)$ with 
unit vectors $(\hat{x},\hat{y},\hat{z})$
by solving the following radiation magnetohydrodynamic (MHD) 
equations \citep[e.g.,][]{Jiangetal2012}
\begin{eqnarray}\label{eqn:equations}
\frac{\partial\rho}{\partial t}+\bfnabla\cdot(\rho \bv)&=&0, \nonumber \\
\frac{\partial( \rho\bv)}{\partial t}+\bfnabla\cdot({\rho \bv\bv-\bb\bb+P^{\ast}{\sf I}}) &=&-\bm{ S_r}(\bP)- \rho g\bm{ \hat z},\  \nonumber \\
\frac{\partial{E}}{\partial t}+\bfnabla\cdot\left[(E+P^{\ast})\bv-\bb\left(\bb\cdot\bv\right)\right]&=&-cS_r(E)-g\rho\bv\cdot\bm{\hat z},  \nonumber \\
\frac{\partial \bb}{\partial t}-\bfnabla\times\left(\bv\times\bb\right)&=&0,\nonumber \\
\frac{\partial E_r}{\partial t}+\bfnabla\cdot \bF_r&=&cS_r(E), \nonumber \\
\frac{1}{c^2}\frac{\partial \bF_r}{\partial t}+\bfnabla\cdot{\sf P}_r&=&{\bf S_r}(\bP),
\end{eqnarray}
where the radiation source terms are
\begin{eqnarray}\label{eqn:sources}
{\bf S_r}(\bP)&=&-\rho\left(\kappa_{\rm aF}+\kappa_{\rm sF}\right)\left[\bF_r-\left(\bv E_r+\bv\cdot{\sf P}_r\right)\right]/c \nonumber \\
&+&\rho\bv(\kappa_{\rm aP}a_rT^4-\kappa_{\rm aE}E_r)/c,\nonumber\\
S_r(E)&=&\rho(\kappa_{\rm aP}a_rT^4-\kappa_{\rm aE}E_r) \nonumber \\
&+&\rho(\kappa_{\rm aF}-\kappa_{\rm sF})\frac{\bv}{c^2}\cdot\left[\bF_r-\left(\bv E_r+\bv\cdot{\sf  P}_r\right)\right].
\end{eqnarray}
Here $\rho,P,\bv,c$ are the gas density, pressure,
flow velocity and speed of light respectively while $P^{\ast}\equiv P+B^2/2$ is the sum of 
gas pressure and magnetic pressure. Notice that the unit of magnetic field is chosen such that 
magnetic permeability is $1$.  The total gas energy
density is $E=E_g+\rho v^2/2+B^2/2$, where $E_g=P/(\gamma-1)$ is the
internal gas energy density with a constant adiabatic index
$\gamma=5/3$.  The gas pressure is $P=\rho \kb T/\mu$, where
$\kb$ is Boltzmann's constant and $\mu=0.62m_p$
is the mean molecular weight for nearly fully ionized gas with proton 
mass $m_p$.  The
radiation constant is $a_r=7.57\times10^{15}$ erg cm$^{-3}$ K$^{-4}$,
while $E_r, \bF_r$ are the radiation energy density and flux.  The
Rosseland mean absorption and scattering opacities are denoted by 
$\kappa_{aF}$ and $\kappa_{sF}$, while $\kappa_{aP}$ and $\kappa_{aE}$ are the Planck
and energy mean absorption opacities. 
We use  a variable Eddington tensor (VET), which is calculated by solving 
the time-independent radiation transfer equation based on the short-characteristic 
method \citep[][]{Davisetal2012}, to relate the radiation pressure ${\sf P}_r$ 
with the radiation energy density $E_r$ such that the radiation moment 
equations are closed. We solve these equations
using the same radiation MHD code {\sc Athena} as in \cite{Jiangetal2015} 
with the MHD module turned on. The code is described and
tested in \cite{Jiangetal2012} with additional improvements
described in \cite{Jiangetal2013b}.

\subsection{Model Parameters}
\label{sec:model}
\cite{Jiangetal2015} show that the important parameters to distinguish 
different regimes of convection are optical depth per pressure scale 
height at the iron opacity peak $\tau_0$, as well as the 
ratio between the speed of light and isothermal sound speed $\tau_c$.
When $\tau_0\gg \tau_c$ as in the post main sequence giant stars, convection is efficient 
and close to the adiabatic convection regime with adiabatic index $4/3$. When 
$\tau_0\ll \tau_c$ as in the main sequence massive stars, convection is inefficient because 
of rapid diffusion and porosity effects 
become significant. This is also the regime where accurate radiation 
transfer is crucial.  Because the iron opacity peak is closer to the surface in this case, 
the magnetic field amplified by  convection could easily reach 
the photosphere and affect the observable properties of massive stars.

Therefore we focus on the parameter regime corresponding to 
the run {\sf StarTop} shown in Table 1 of \cite{Jiangetal2015} 
with $\tau_0=166.5$ and $\tau_c=6.54\times 10^3$. The model parameters are taken from a 
\mesa \citep{Paxtonetal2011,Paxtonetal2013,Paxton:2015} 
calculation of a main sequence
$80\msun$ star. The simulation box is located 
at radius $13.6R_{\odot}$ with a constant gravitational acceleration $g=1.17\times10^4$ 
cm/s$^2$ and a constant radiation flux $F_{r,i}=3.06\times 10^{14}$ erg/(cm$^2$ s) 
applied at the bottom of the simulation box, which are the same as in {\sf StarTop}. 
Horizontal and vertical sizes of the simulation box are $L_x=L_y=1.92H_0$ and $L_z=5.12H_0$ 
with resolution $N_x=N_y=192$ and $N_z=512$, where $H_0=2.37\times 10^{10}$ cm 
is the fiducial pressure scale height. The fiducial density, temperature, velocity, gas pressure 
and time units are $\rho_0=5.52\times 10^{-9}$ g/cm$^3$, $T_0=1.57\times 10^5$ K, 
$v_0=4.59\times 10^6$ cm/s, $P_0=1.16\times 10^5$ dyn/cm$^3$ and $t_0=1.42\times 10^3$ s. 
These fiducial units are summarized in Table \ref{Table:units}.
The initial gas and radiation pressure at the iron opacity peak is $P_0$ and $13.2P_0$.
Compared with {\sf StarTop}, the horizontal box size is increased 
by $60\%$ with a comparable resolution.

The total Rosseland mean opacity $\kappa_t\equiv \kappa_{\rm aF}+\kappa_{\rm sF}$ 
in the simulation is calculated based on opacity tables taken from \mesa with 
assumed metallicity $Z=0.02$ and hydrogen fraction $X=0.6$ (Figure 2 of \citealt{Jiangetal2015}). 
We assume a constant 
electron scattering opacity $\kappa_{\rm sF}=0.32$ cm$^2$/g and absorption opacity is simply 
calculated as $\kappa_{\rm aF}=\kappa_t-\kappa_{\rm sF}$. The Planck and energy mean absorption 
opacities $\kappa_{\rm aP},\ \kappa_{\rm aE}$ are taken to be the same as $\kappa_{\rm aF}$.

\begin{table}[h]
\caption{The Fiducial Units}
\begin{center}
\begin{tabular}{cc}
\hline
Variables/Units & Values\\
\hline
$H_0$ / cm &  $2.37\times 10^{10}$ \\
$\rho$ / g cm$^{-3}$  &  $5.52\times 10^{-9}$\\
$T_0$ / K  &  $1.57\times 10^5$ \\
$v_0$ / cm s$^{-1}$ & $4.59\times 10^6$ \\
$P_0$ / dyn cm$^{-3}$ & $1.16\times 10^5$\\
$t_0$ / s   &  $1.42\times 10^3$\\
\hline
\end{tabular}
\end{center}
\label{Table:units}
\end{table}

\subsection{Simulation Setup}
Initial vertical profiles of the gas and radiation quantities are constructed in the same way 
as described in Section 3.3 of \cite{Jiangetal2015}, which are calculated 
based on equations of hydrostatic and thermal 
equilibrium. We add a uniform vertical ($B_{z,0}$) and horizontal magnetic 
field ($B_{y,0}$) through the whole simulation box, which does not affect the initial conditions of 
other quantities. Values of magnetic field strengths and configurations we have 
explored are listed in Table \ref{Table:parameters}.  For simulations {\sf StarB1}, 
{\sf StarB2} and {\sf StarB3}, the magnetic field is inclined by 
$26.4^{\circ}$ with respect to the horizontal plane but they have different magnetic field 
strengths. Simulation {\sf StarB4} has the same $B_{z,0}$ as {\sf StarB3} 
but $B_{y,0}$ is increased by a factor of 5 compared with 
{\sf StarB3}. Therefore, 
the magnetic field has an initial inclination angle $5.7^{\circ}$ with respect to the horizontal plane.

\begin{table}[h]
\caption{Initial Magnetic Field Parameters}
\begin{center}
\begin{tabular}{ccccc}
\hline
Variables/Units & {\sf StarB1} & {\sf StarB2} & {\sf StarB3} & {\sf StarB4}\\
\hline
$B_{z,0}/G$              		      &	  60	    & 	 191		& 	382		&	382	\\	
$B_{y,0}/G$			      &	  121	    &   382		&	764		&	3819	\\
$\left(P_g+P_r\right)/P_{B,z}$ & 11373  &	 1137	&	284		&	284	\\
$P_g/P_{B,z}$		      &  800	    &   80		&	20		& 20  	\\
\hline
\end{tabular}
\end{center}
\label{Table:parameters}
\begin{tablenotes}
\item Note: The gas pressure $P_g$ and radiation pressure $P_r$ are the initial values at the 
iron opacity peak. The vertical and horizontal magnetic field components $B_{z,0}$ (with magnetic 
pressure $P_{B,z}$) and $B_{y,0}$ are the initial values and only $B_{z,0}$ is conserved during the simulations. 
Notice that magnetic field unit is converted to Gauss here from the code unit by matching the dimensionless 
ratio $P_g/P_{B,z}$.
\end{tablenotes}
\end{table}

We use the same boundary conditions for the gas and radiation quantities as described in section 3.3 
of \cite{Jiangetal2015}, namely, reflecting boundary condition at the bottom and outflow 
boundary condition at the top. For magnetic field at the bottom boundary, the vertical component $B_z$ is 
copied from the last active zones to the ghost zones while the horizontal components $B_x$ and 
$B_y$ are copied with the opposite sign. We have also tried to just copy $B_x$ and $B_y$ from the last 
active zones to the ghost zones, which does not make any noticeable difference. 
At the top boundary, the vertical component $B_z$ 
is copied from the last active zones to the ghost zones and the horizontal components $B_x$ 
and $B_y$ are also copied when the vertical component of flow velocity $v_z$ points outward. 
When $v_z$ points inward, $B_x$ and $B_y$ are set to be zero in the ghost zones to avoid magnetic field 
being carried into the simulation domain from the top boundary. In this setup, the volume integrated $B_z$ 
is a conserved quantity while the volume integrated $B_x$ and $B_y$ can change. In particular, 
$B_x$ and $B_y$ can escape through the open top boundary while $B_z$ cannot. 

\section{Results}
\label{sec:result}
In this section, we describe how the initial development of convection 
and the final turbulent structures of the envelope are modified by 
different magnetic field strengths and configurations. For any 
quantity $a$, we will use $\langle a\rangle$ to represent the horizontal averaged value 
for each height $z$.

\subsection{Linear Growth Phase of the Instabilities}

\begin{figure}[h]
\begin{center}
\includegraphics[width=1.0\columnwidth]{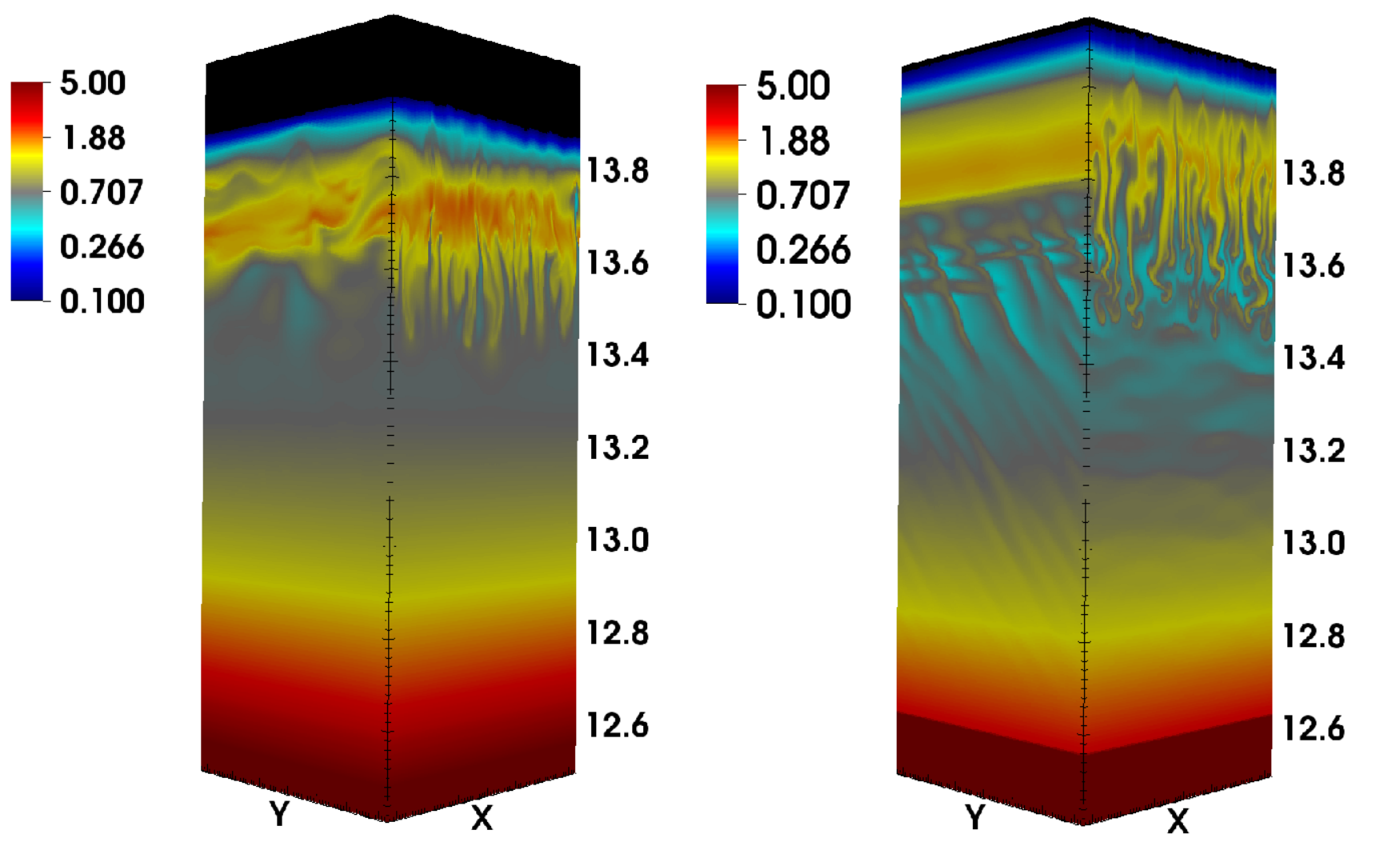}
\caption{Initial development of the magnetically modified convection 
for simulations {\sf StarB3} at time $12.8t_0$ (left) and {\sf StarB4} 
at time $15.0t_0$ (right). Convective plumes are 
visible at the regions of density inversion. The shock trains in the right 
panel are due to photon bubble instability. 
(Movies showing the density evolutions 
of the two runs are available at \url{
 https://goo.gl/3kYbtg}).
}
\label{snapshot_early}%
\end{center}
\end{figure}

\begin{figure*}[htp]
\begin{center}
\includegraphics[width=1.0\columnwidth]{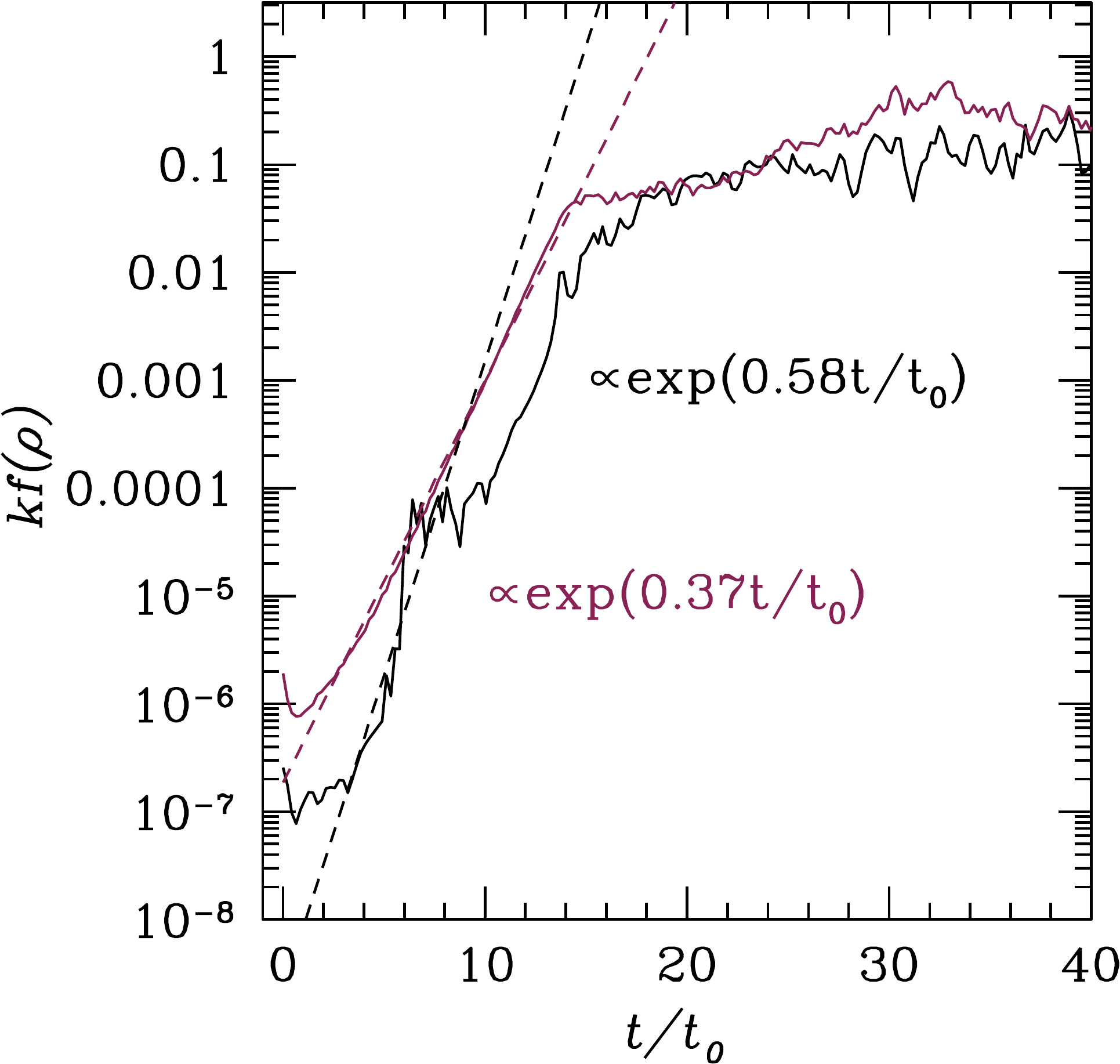}
\includegraphics[width=1.0\columnwidth]{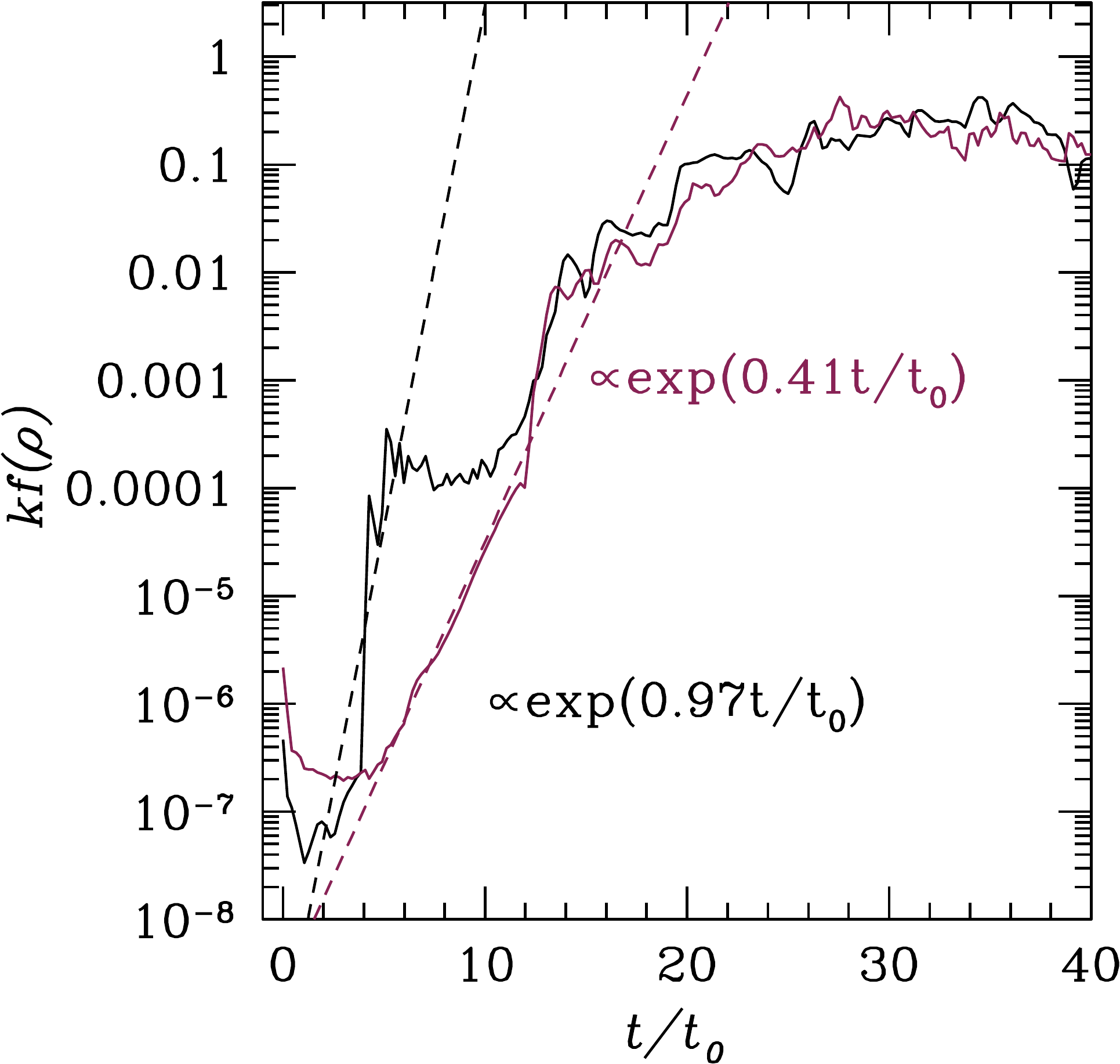}
\caption{History of the density power spectrum $kf(\rho)$ for the 
run {\sf StarB4} at $z=13.6\rsun$ (left) and $z=13.9\rsun$ (right) 
during the first $40t_0$. The solid black and red lines are for the 
modes with wavelengths $H_0$ and $0.13H_0$ while the dashed lines 
indicate the best fitted exponential growth rate for these modes. 
The growth rates in the left and right panels are for the 
photon bubble and Parker instabilities respectively. 
}
\label{growthrate}
\end{center}
\end{figure*}

As in the hydro case, there is a density inversion around $13.8R_{\odot}$ 
in the initial hydrostatic envelope at the iron opacity peak where the local radiation acceleration is 
larger than gravitational acceleration. The initial profile we construct is actually 
pretty similar to the 1D \mesa profile in this regime, as convection 
is inefficient to transport energy. In our 3D simulations, the density inversion is unstable to 
convection, resulting in the development of turbulence after $\sim 20-40t_0$. 
In simulation {\sf StarB1} when the initial magnetic pressure is much smaller than 
both the gas and radiation pressure, the initial growth rate and the dominant 
wavelength of the convection are very  similar to the hydro case shown in \cite{Jiangetal2015}. 
For simulation {\sf StarB3} when the initial total magnetic pressure is comparable 
to the gas pressure at the iron opacity peak, it only takes $\sim 20t_0$ to completely destabilize the 
density inversion, compared to $\sim 40t_0$ in {\sf StarB1}. The fastest 
growing mode is also modified by the presence of a magnetic field. 
Figure 16 of \cite{Jiangetal2015} 
shows that during the linear growth phase of convection, the long wavelength 
mode dominates and it is roughly symmetric with respect to $x$ and $y$ directions. 
This is also true for run {\sf StarB1} when the initial magnetic field is weak. 
However, during the linear growth phase of {\sf StarB3}, as shown in the left panel 
of Figure \ref{snapshot_early}, the magnetic field breaks the symmetry. 
Short wavelength modes dominate along the direction perpendicular 
to the magnetic field,  and are the result of Parker instability. For the run {\sf StarB4} 
when $B_y$ is increased such that the angle between the magnetic field and the horizontal 
direction is decreased, we see both the instability associated with the density 
inversion as well as shock trains in the envelope as shown in 
the right panel of Figure \ref{snapshot_early}.
This is very similar to the nonlinear outcome of the photon bubble
instability \citep[][]{Begelman2001,Turneretal2005}. 
The shock fronts are inclined by an angle of roughly 
$40^{\circ}$ with respect to the direction of gravity.

The properties of these instabilities compare well with linear theory.
For the photon bubble instability, the maximum growth rate is achieved for
wavelengths shorter than $c_{g}^2/g$ \citep[][]{BlaesSocrates2003}, where 
$c_g$ is the gas sound speed.  However, this very short length scale is
unresolved by the simulation, and is in fact optically thin so the diffusive
transport of photons that drives the instability does not even apply at these
wavelengths.  Longer wavelength growth rates can be estimated by neglecting
the sound speed in the gas alone \citep[][]{Gammie1998,BlaesSocrates2003},
giving a dispersion relation for the angular frequency $\omega$ of the mode
\begin{equation}
\omega^2=igk(\bm{\hat k}\cdot\bm{\hat b})[(\bm{\hat k}\cdot\bm{\hat b})
(\bm{\hat k}\cdot\bm{\hat z})\Theta_\rho-(\bm{\hat k}\times\bm{\hat b})\cdot
(\bm{\hat k}\times\bm{\hat z})].
\end{equation}
Here $k\bm{\hat k}$ is the wavevector of the wave and $\bm{\hat b}$ is a unit
vector in the direction of the magnetic field.  The quantity $\Theta_\rho$ is
the logarithmic derivative of the Rosseland mean opacity with respect to density
at fixed temperature, and is approximately 0.1 throughout our simulation, and
can therefore be neglected.  This then gives a photon bubble growth rate that
depends only on wavelength and the direction of propagation, and is almost
independent of height.  For {\sf StarB4}, the magnetic field is inclined by
$6^\circ$ to the horizontal, and this gives a maximum growth rate for wavefronts
that are inclined by $40^\circ$ to the vertical, exactly as we observe in
the simulations.  Inclining the magnetic field further away from the
horizontal reduces the maximum growth rate, which may explain why there are
no obvious shock trains in the other simulations.

In order to estimate the 
linear growth rates in the simulations,  
we perform a Fourier transform of density at 
heights $z=13.6R_{\odot}$ and $z=13.9R_{\odot}$ for each snapshots.  Histories 
of the binned 1D power spectrums for two different wave numbers are shown 
in Figure \ref{growthrate} at $z=13.6R_{\odot}$ (left panel) and $z=13.9R_{\odot}$ 
(right panel). The growth rates are larger at higher height because the horizontal
magnetic field can leave the simulation box from the top boundary, which results in a
larger magnetic pressure gradient at $z=13.9R_{\odot}$.  The maximum photon bubble growth
rate for a wavelength equal to $H_0$ is $0.66/t_0$, and this compares favorably to
the measured growth rate of $0.58/t_0$ at $z=13.6\rsun$. 
At $z=13.9\rsun$, the medium is actually supported by magnetic pressure
gradients, and this is Parker unstable.  While photon bubbles and Parker instabilities
arise from the same underlying mode \citep[][]{TaoBlaes2011}, 
Parker dominates for short
horizontal wavelengths that are perpendicular to the magnetic field, and this is exactly
what we see in the simulations in this region.  When photons diffuse rapidly and the
gas pressure is negligible, as is the case here, the maximum instability growth rate
is $[g/2H_{\rm mag}]^{1/2}$, where $H_{\rm mag}$ is the local magnetic pressure scale
height \citep[][]{TaoBlaes2011}.  This gives a peak growth rate of approximately $1/t_0$
peaking in the region $z=13.9-14.0\rsun$, independent of wavelength, in rough agreement
with the growth rates we find in this region. Notice that these growth rates are 
much larger than the growth rate of hydrodynamic convection, which is $0.16/t_0$ for wavelength 
$H_0$.
The smaller growth rates for the short wavelength 
modes at both heights in Figure \ref{growthrate} are probably because of extra numerical dissipation, which 
we also observe in the run {\sf StarTop} \citep[][]{Jiangetal2015}.

\subsection{Steady State Turbulent Structures}

\begin{figure}[h]
\begin{center}
\includegraphics[width=1.0\columnwidth]{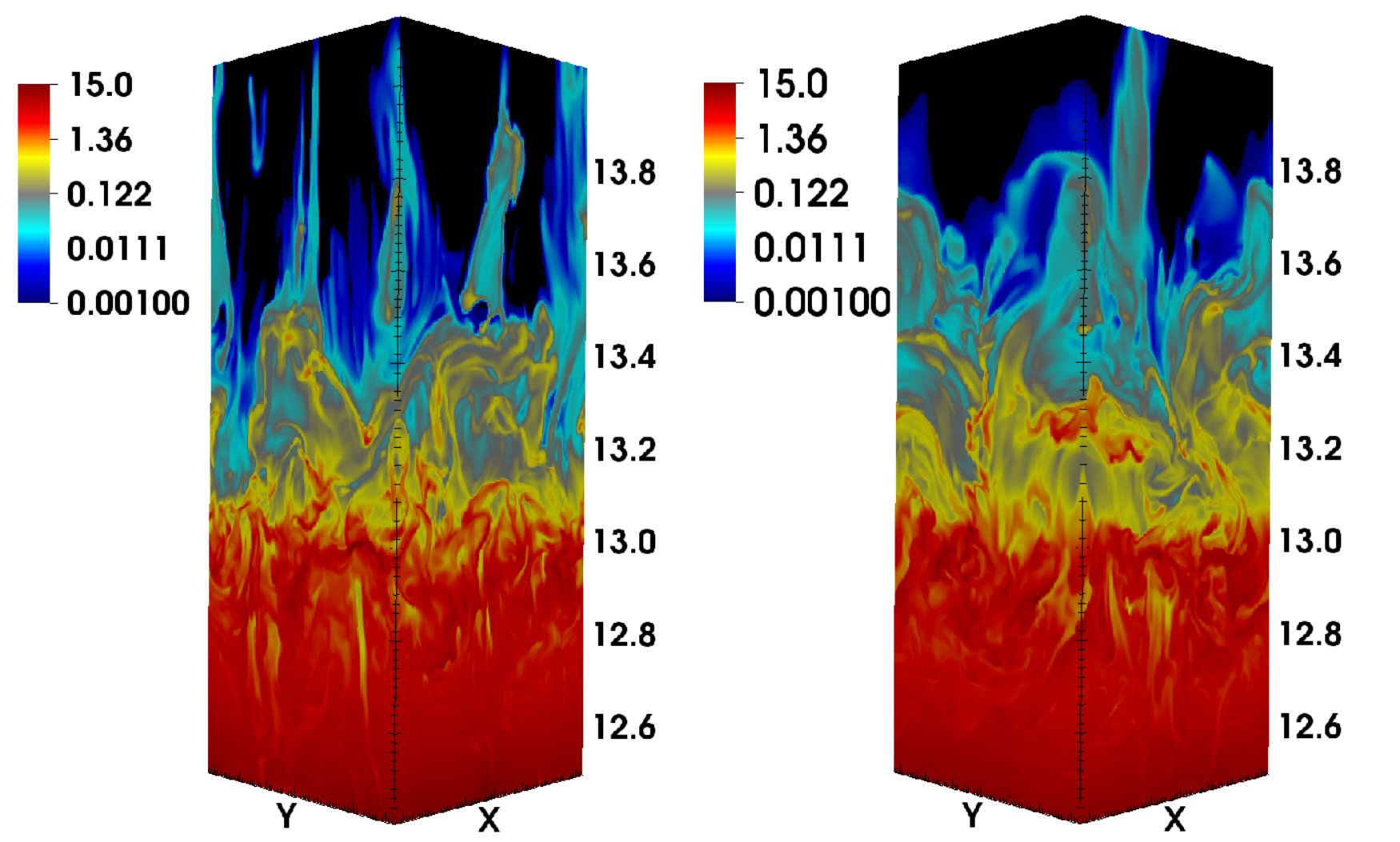}
\caption{Final turbulent states of the envelope for 
simulations {\sf StarB3} at time $94.0t_0$ (left) and 
{\sf StarB4} at time $68.4t_0$ (right). Despite different 
initial evolutions as shown in Figure \ref{snapshot_early}, the final 
states of the two simulations 
are very similar because the horizontal magnetic fields are lost. 
\label{snapshot}%
}
\end{center}
\end{figure}

\begin{figure}[htp]
\begin{center}
\includegraphics[width=1.0\columnwidth]{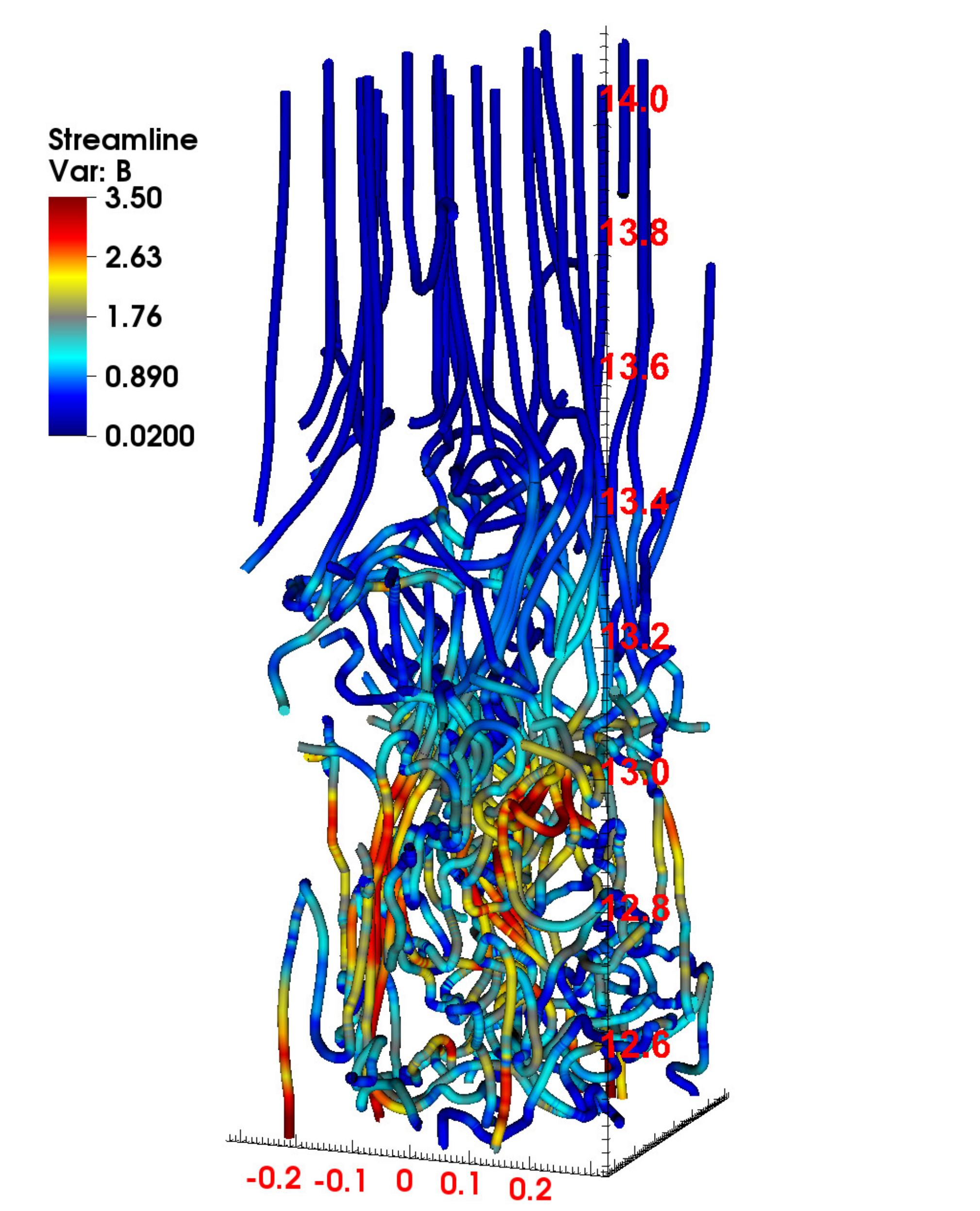}
\caption{Snapshot of magnetic field streamline is at time $94.0t_0$ for simulation {\sf StarB3}. 
Color of the streamlines is the magnetic field strength in unit of $\sqrt{4\pi P_0}$. 
The tangled magnetic fields below $z=13.4R_{\odot}$ are due to convection. Above 
this region, magnetic fields are dominated by the vertical component. 
\label{Bline}%
}
\end{center}
\end{figure}

Although the shock trains caused by the photon bubble instability show up in run 
{\sf StarB4} but not {\sf StarB3}, the final turbulent structures of the envelopes in 
the two runs are very similar. Snapshots of density for the two runs after the initial 
development of the instabilities are shown in 
Figure \ref{snapshot}. The entire envelopes are turbulent due to 
convection, while the shock trains from the photon bubble instability 
are hardly noticeable in {\sf StarB4}. The Parker instability has grown to
completely dominate the turbulent state, and in fact the magnetic field
geometry has completely changed from its initial configuration.  As shown
in Figure~\ref{Bline}, the magnetic field is dominated by the vertical
component, as the horizontal component has buoyantly left the simulation domain.
This vertical field configuration is also less able to support the weight
of high density regions within photon bubble shock trains (cf. the right hand
panel of Figure 1 of \citealt{Turneretal2005}).
Figure \ref{Bline} also shows that a small scale 
turbulent magnetic field dominates in the iron opacity region. Above $13.6R_{\odot}$, 
only the vertical magnetic field survives.


\begin{figure*}[htp]
\begin{center}
\includegraphics[width=1.0\columnwidth]{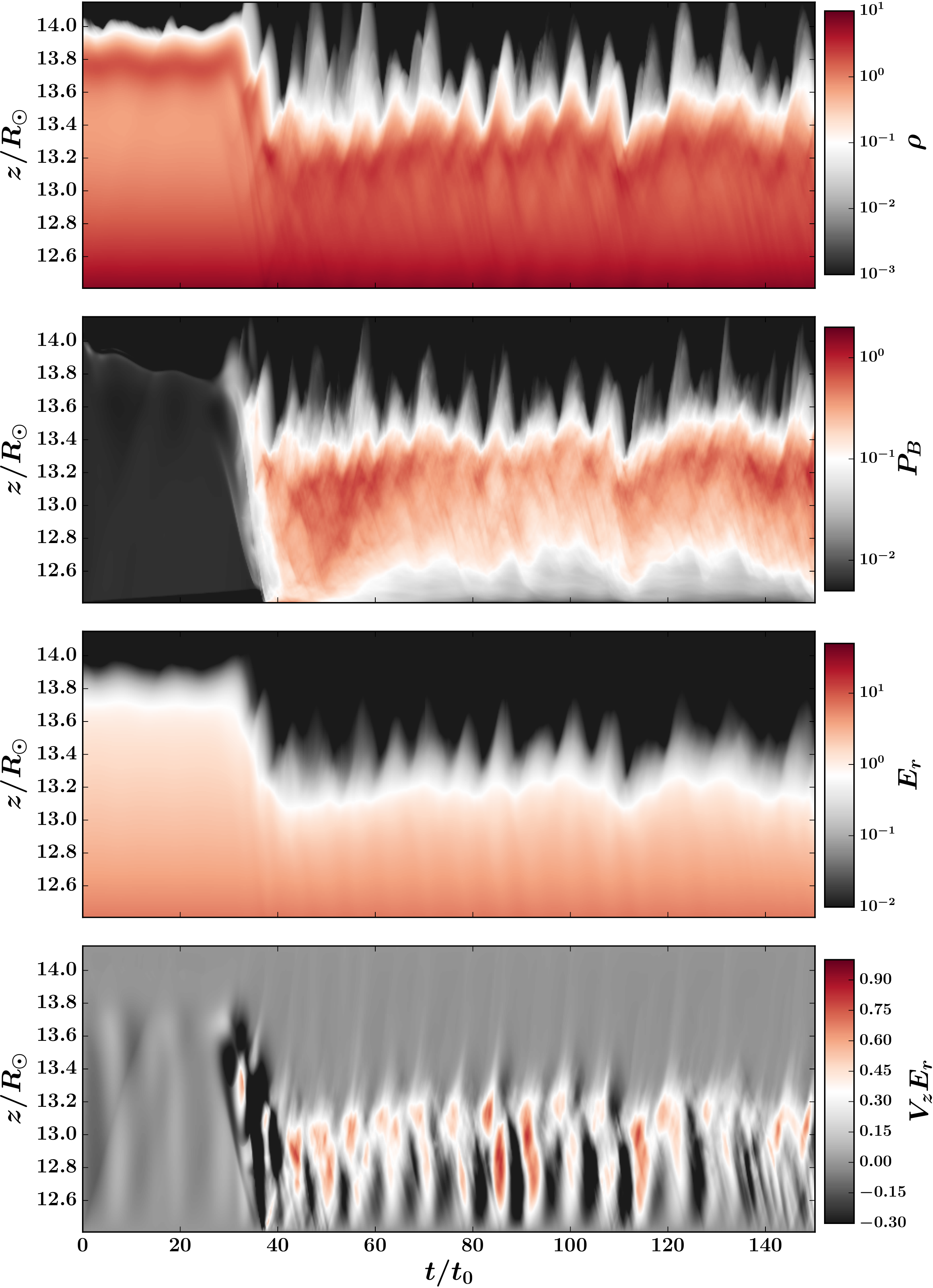}
\includegraphics[width=1.0\columnwidth]{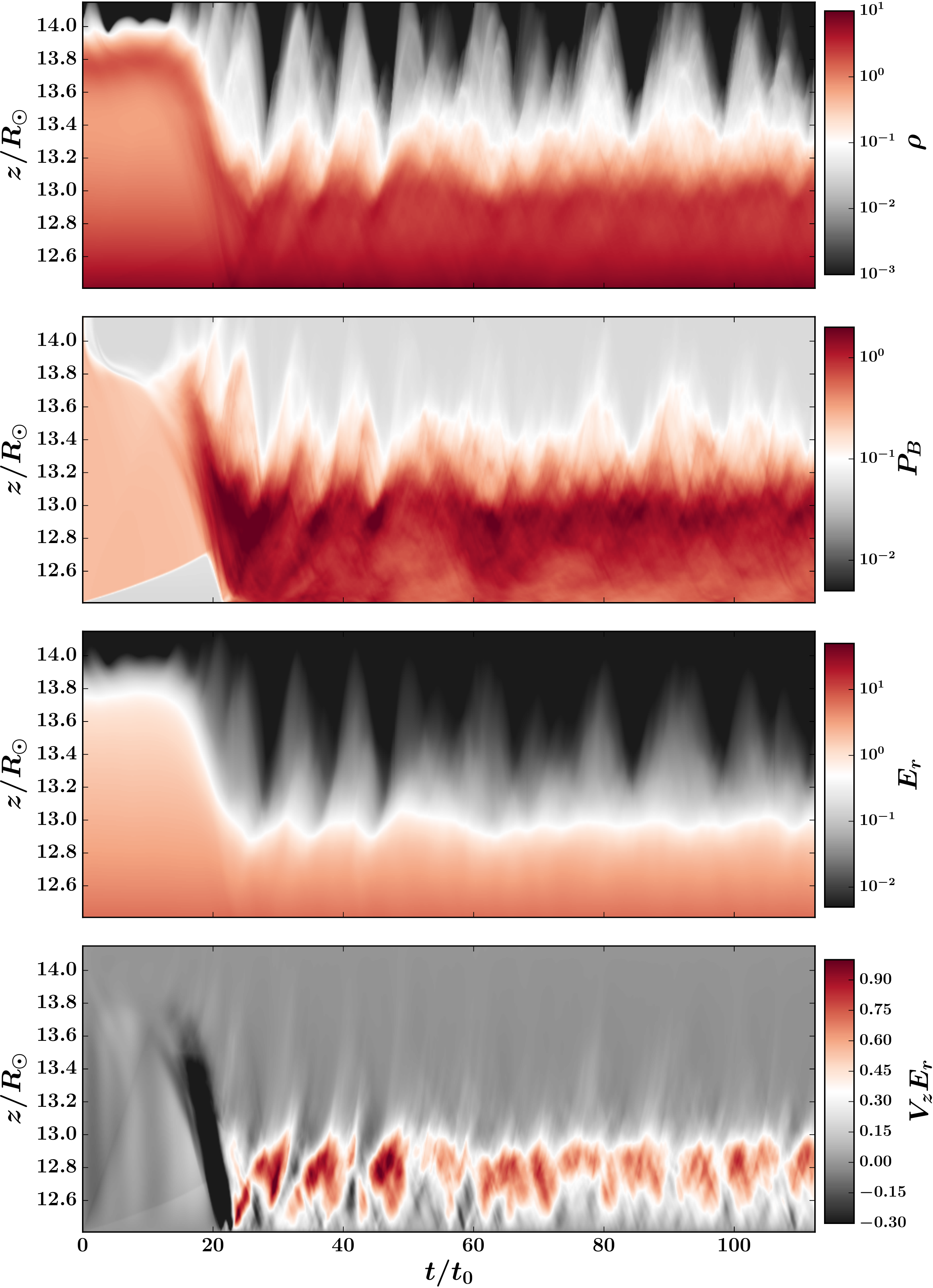}
\caption{Time evolutions of the horizontally averaged vertical 
profiles of density $\rho$, magnetic pressure $P_B$, radiation 
energy density $E_r$ and advective radiation flux $V_zE_r$ from  
simulations {\sf StarB1} (the left panel) and {\sf StarB3} (the right panel). 
Units of these quantities are 
$\rho_0, P_0,a_rT_0^4,v_0a_rT_0^4$ as given in section \ref{sec:model}. 
After the initial development of instabilities ($\sim40t_0$ for {\sf StarB1} 
and $\sim 20t_0$ for {\sf StarB3}), the envelopes reach steady states with 
regular vertical oscillations. 
\label{History}%
}
\end{center}
\end{figure*}

The whole histories of the horizontally averaged vertical profiles of $\rho$, $P_B$, 
$E_r$ and $v_zE_r$ for simulations {\sf StarB1} and {\sf StarB3} are shown in 
Figure \ref{History}. 
After the onset of turbulence, the stellar envelope undergoes regular 
vertical oscillations as shown in Figure \ref{History}, which is also observed 
in the pure hydrodynamic simulation. The oscillation amplitude increases with 
increasing magnetic field strength, which causes larger temporal variability at the photosphere.
Vertical advection flux $V_zE_r$ oscillates with the envelope. However, 
there is always more outward advection flux than inward advection flux, which 
results in net positive advection flux.  

\begin{figure*}[htp]
\begin{center}
\includegraphics[width=1.0\columnwidth]{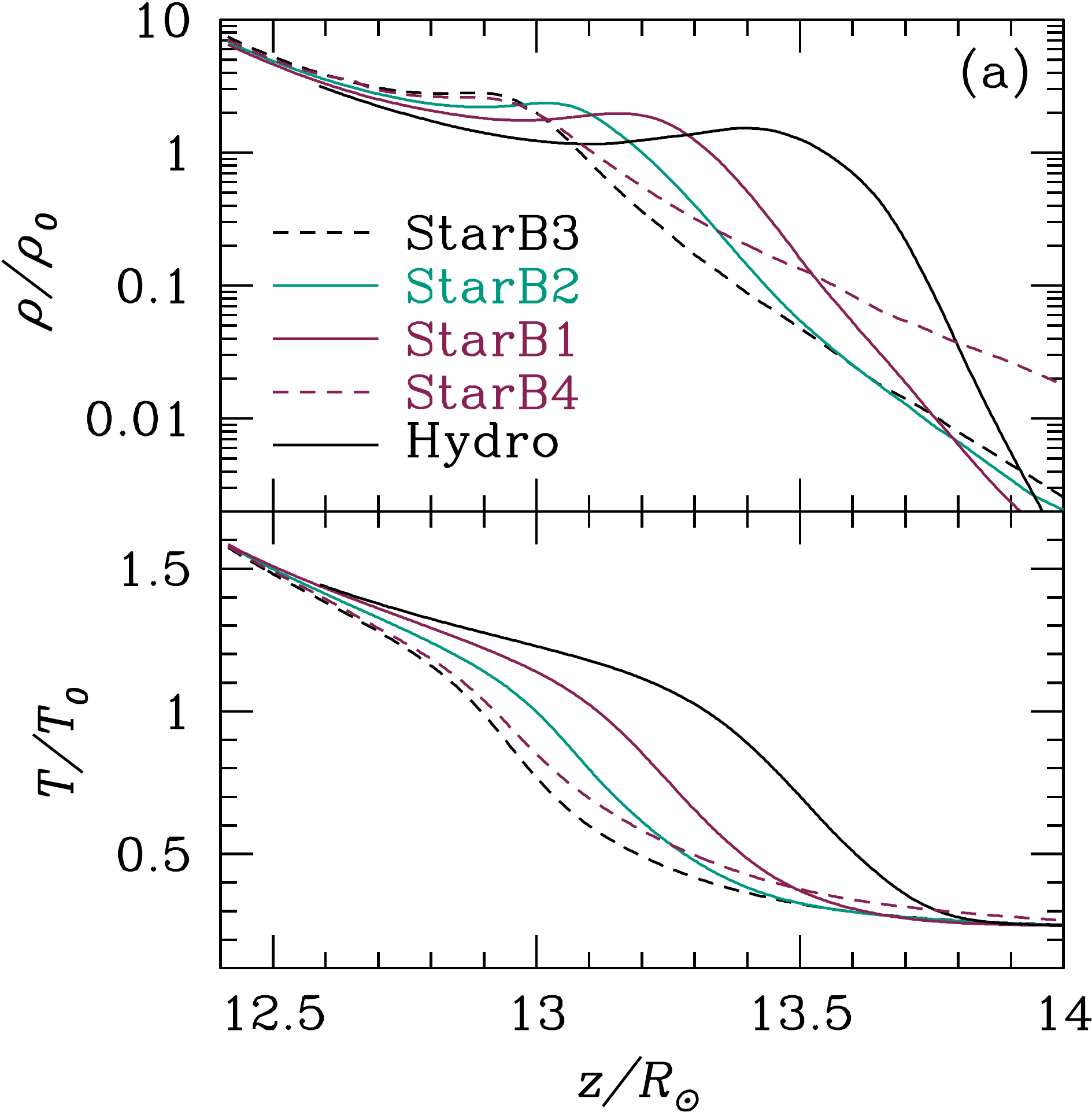}
\includegraphics[width=1.0\columnwidth]{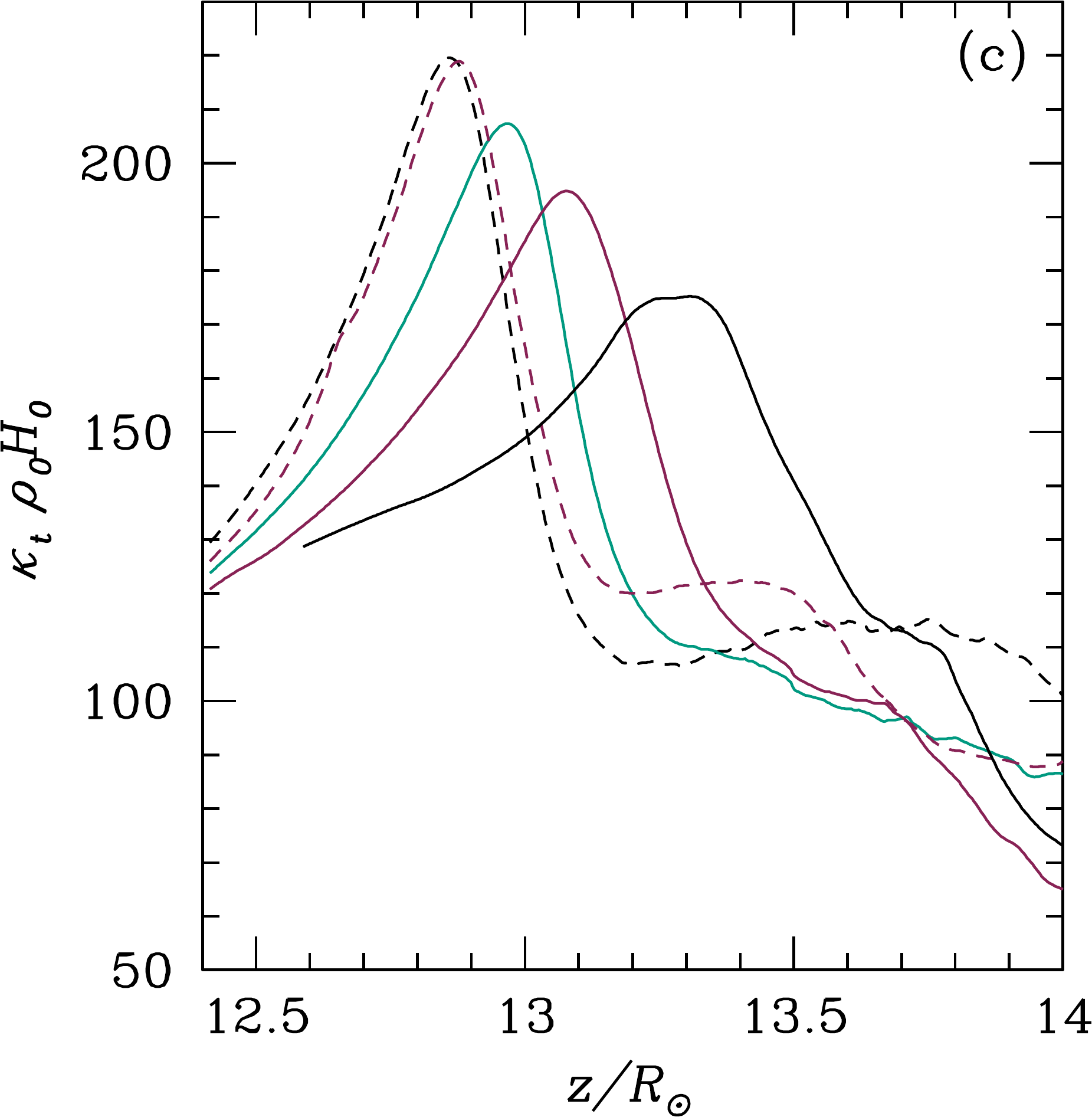}\\
\includegraphics[width=1.0\columnwidth]{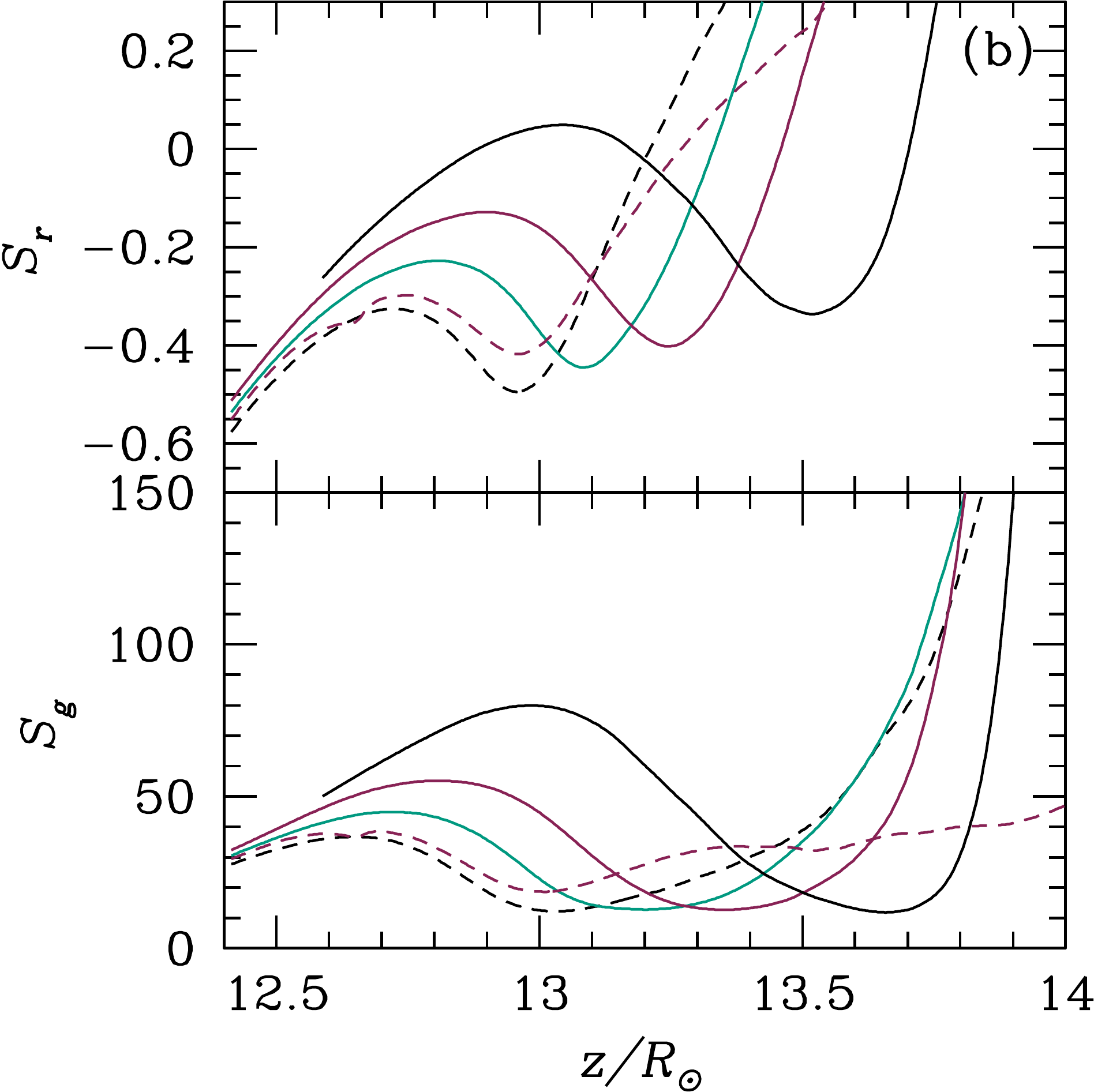}
\includegraphics[width=1.0\columnwidth]{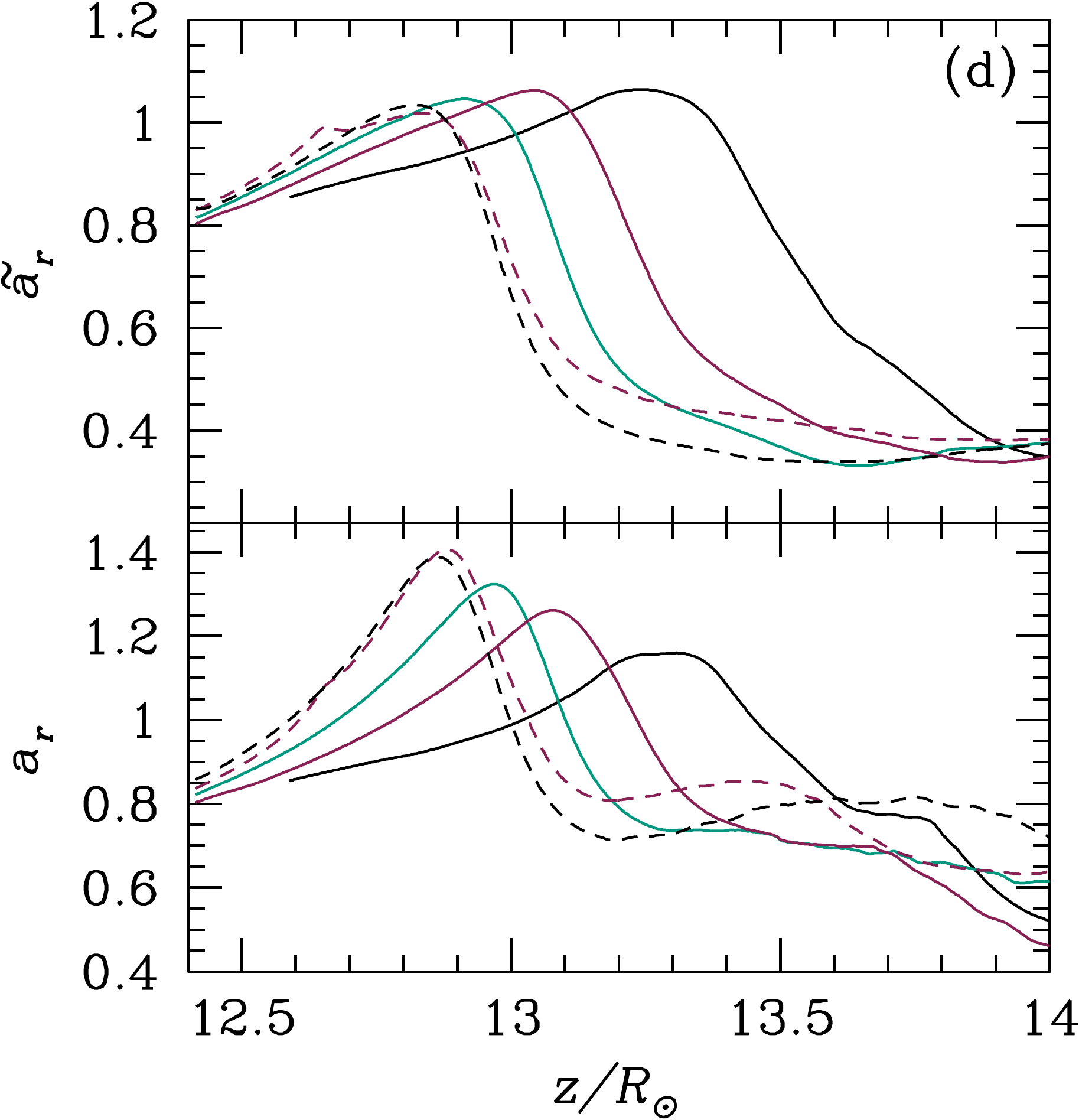}
\caption{The time and horizontally averaged vertical profiles of density (top panel of a), 
temperature (bottom panel of a), 
entropy (panel b), opacity (panel c) as well as radiation accelerations (panel d) for the 
run {\sf StarTop} without magnetic fields as well as the four runs with magnetic fields 
{\sf StarB1,\ StarB2,\ StarB3,\ StarB4}. The unit of the entropy is $\kb/\mu$ while 
the radiation accelerations $a_r$ and $\tilde{a}_r$ are scaled with the gravitational 
acceleration.The whole envelopes shrink to smaller heights with increasing $B_z$ 
due to larger porosity factors.%
}
\label{Profile}
\end{center}
\end{figure*}

It is clear from Figure \ref{History} that the averaged positions 
of the oscillated envelopes and the oscillation amplitude are 
different with different magnetic field strengths. The time and 
horizontally averaged vertical profiles of density, opacity, entropy 
and radiation accelerations during the oscillation periods 
for the runs {\sf StarTop, StarB1, StarB2, StarB3} and {\sf StarB4} 
are shown in Figure \ref{Profile}. Compared with the hydro case, 
the whole envelopes shrink and the iron opacity peaks move 
from $z=13.30R_{\odot}$ as in {\sf StarTop} to $13.08R_{\odot}$ 
in {\sf StarB1}, $12.97R_{\odot}$ in {\sf StarB2}, $12.86R_{\odot}$ 
in {\sf StarB3}. The change of iron opacity peak locations from 
{\sf StarTop} to {\sf StarB3} corresponds to $1.3H_0$. The 
three runs {\sf StarB1, StarB2, StarB3} also show that for 
the same initial magnetic field configuration, larger initial magnetic fields strengths correspond to 
larger envelope shrinking. 
The field geometry does not seem to play an important role for the final envelope structure, as shown by the fact that
 {\sf StarB3} and {\sf StarB4} (same initial amplitude, different geometry) 
reach very similar vertical structures, particularly in the 
regions around and below the iron opacity peaks. 
In the regions above the iron opacity peak, the magnetic field plays a significant role in providing support against
gravity, particularly in runs {\sf StarB3} and {\sf StarB4}, which causes the density 
to drop slowly with height. Stronger initial horizontal magnetic field causes stronger 
magnetic pressure support in this region. This is mainly because most of the initial 
horizontal magnetic field has moved to the top 
part of the box due to buoyancy. In the steady state, the buoyantly rising 
magnetic field from the convective region is balanced by the magnetic flux loss through 
the top of the simulation domain. 

When the iron opacity peak is moved to the higher density region, the maximum opacity $\kappa_t$ is 
increased as shown in the panel (c) of Figure \ref{Profile}. For the same radiation flux, the radiation 
acceleration $a_r$ is also increased as shown in the panel (d) of Figure \ref{Profile}. However, 
the density weighted radiation acceleration $\tilde{a}_r$ is actually smaller with stronger magnetic field. 
Therefore, the remaining density inversions in the time averaged structures are weaker with flatter entropy 
profiles as shown in panel (b) of the same Figure. This will be explained in Section \ref{sec:porosity}.

\begin{figure*}[htp]
\begin{center}
\includegraphics[width=1.0\columnwidth]{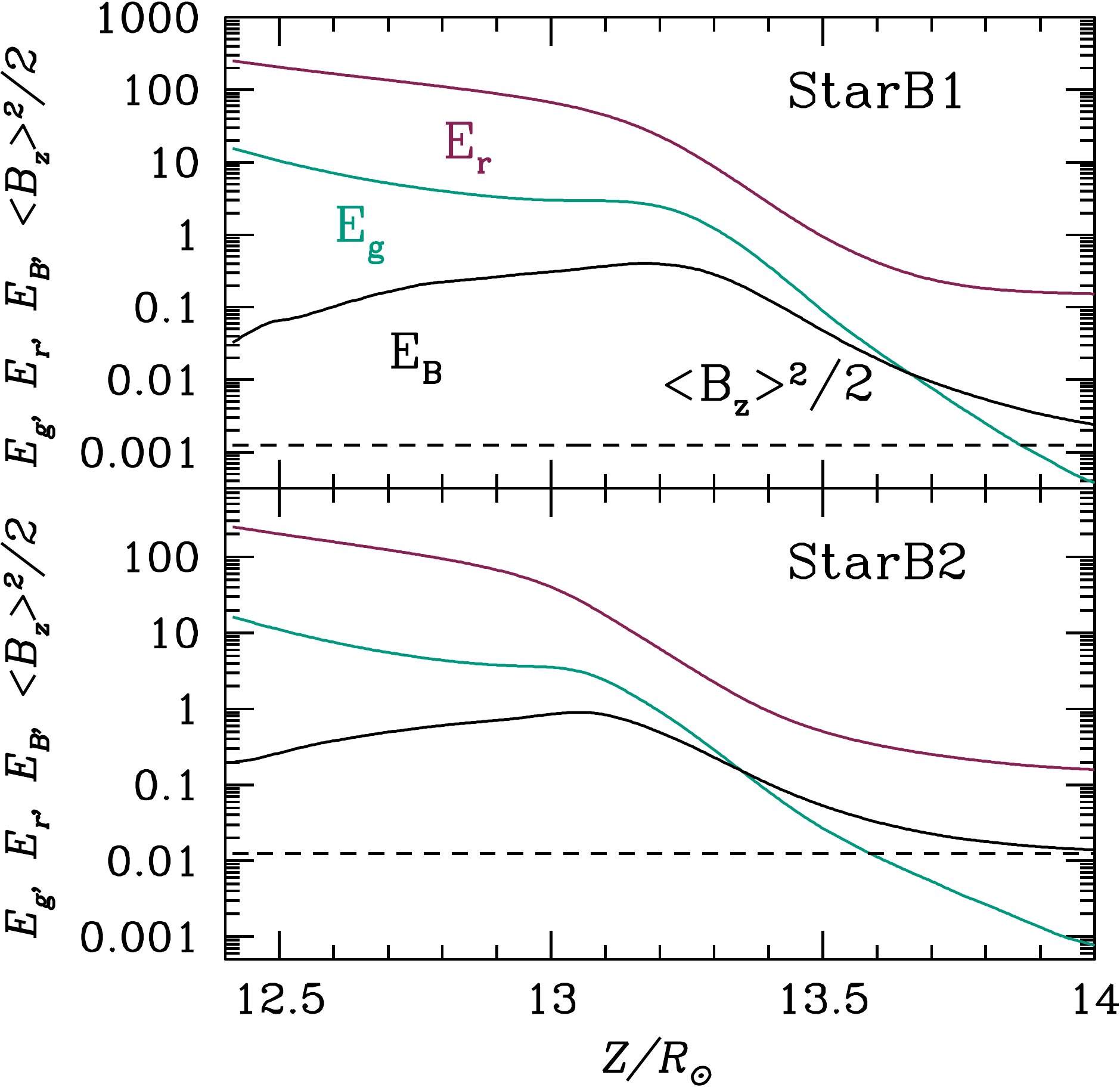}
\includegraphics[width=1.0\columnwidth]{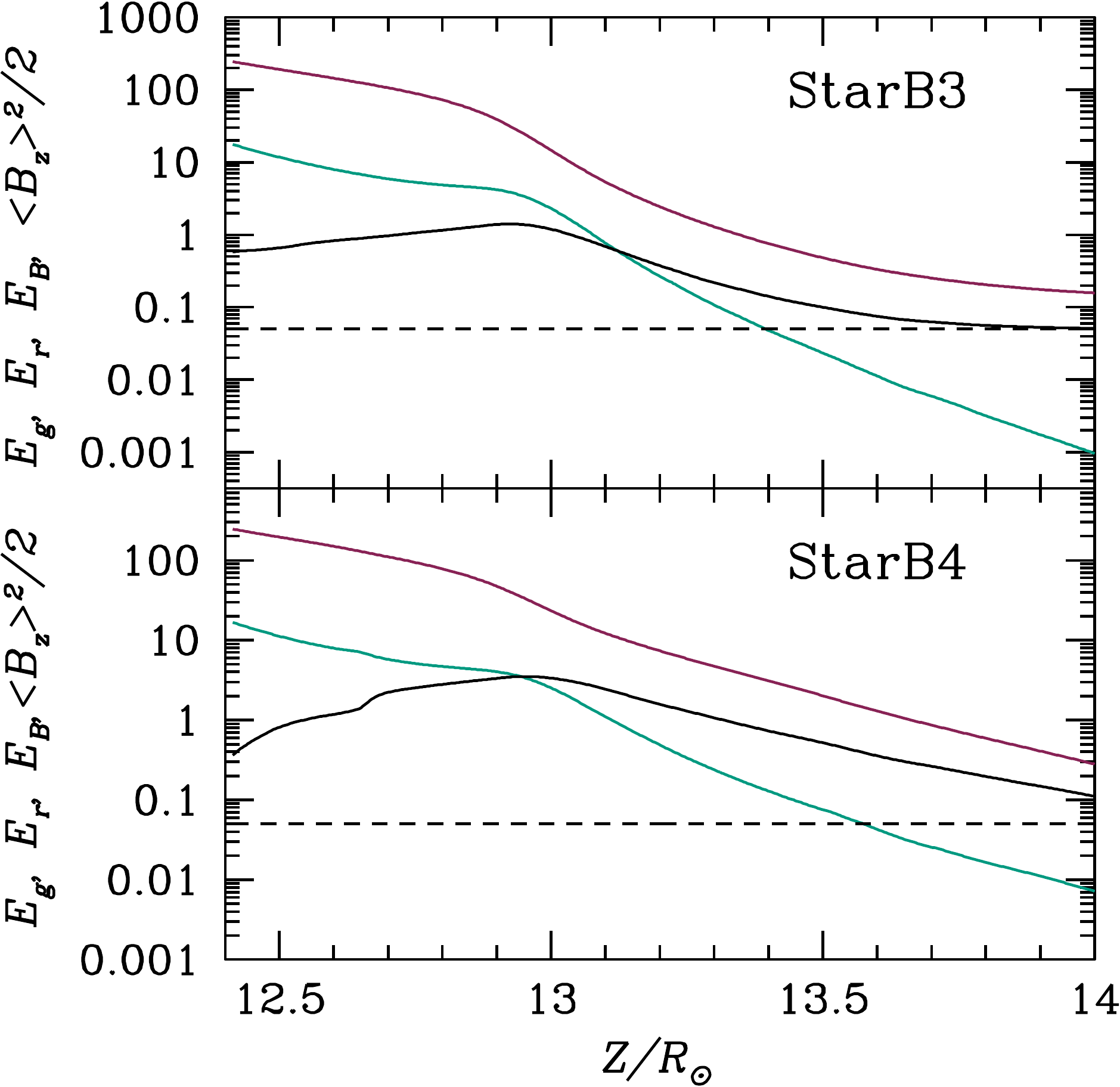}
\caption{Vertical profiles of the time and horizontally averaged 
gas internal energy $E_g$ (the green lines), radiation energy density $E_r$ (the red lines), 
magnetic energy density $E_B$ (the solid black liens) as well as the magnetic energy density associated with the 
horizontally averaged 
mean vertical magnetic field $\langle B_z\rangle^2/2$ (the dashed black lines) for the four runs {\sf StarB1,\ StarB2,\ StarB3.\ StarB4}. 
Unit of the energy densities is $P_0$.
}
\label{energy}
\end{center}
\end{figure*}

The time averaged vertical profiles of radiation energy density $E_r$, gas internal energy density $E_g$, total 
magnetic energy density $E_B$ as well as the magnetic energy 
density associated with the horizontally averaged mean vertical component of magnetic field $\bar{B_z}$ 
are shown in Figure \ref{energy}. Because vertical magnetic flux is conserved, $\bar{B_z}$ is the same as the 
initial vertical magnetic field $B_{z,0}$.  The magnetic energy density includes contributions from the turbulent 
magnetic field as well as the horizontally averaged mean magnetic field. 
In the convective region near the iron opacity peak, the turbulent component associated with the convection 
dominates as $E_B$ is much larger than $(\bar{B}_x^2+\bar{B}_y^2+\bar{B}_z^2)/2$. Above the iron 
opacity peak, $E_B$ gets close to $\bar{B}_z^2/2$. This is also consistent 
with the magnetic field streamlines shown in Figure \ref{Bline}.
In steady state, the vertical component of radiation pressure dominates everywhere for the four runs, 
although in the regions above the iron opacity peak, the magnetic pressure gets close to the radiation 
pressure in {\sf StarB3} and {\sf StarB4}.

\begin{figure}[htp]
\begin{center}
\includegraphics[width=1.0\columnwidth]{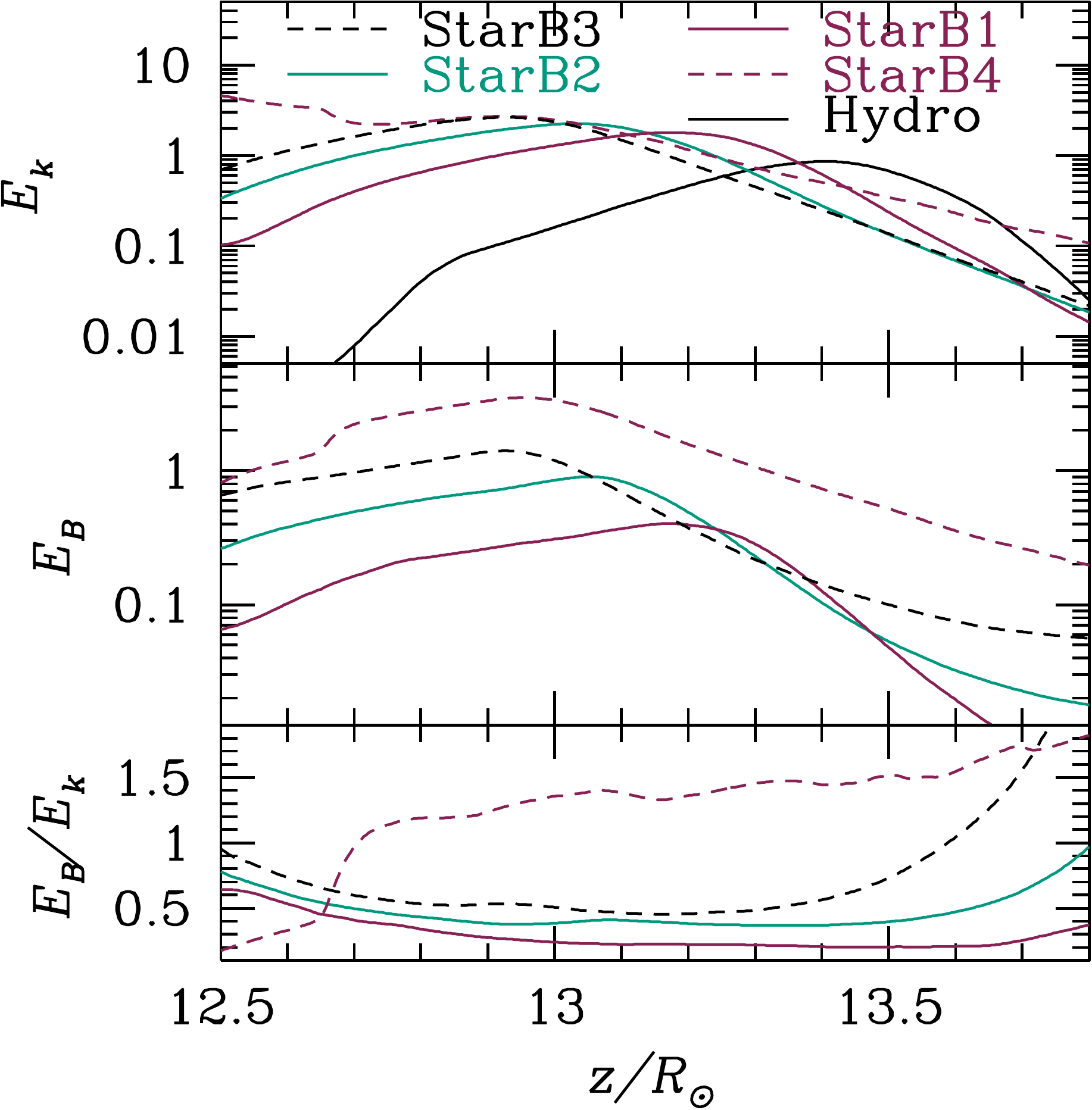}
\caption{Vertical profiles of the time and horizontally averaged kinetic 
energy density (the top panel), energy density due to the vertical 
component of magnetic field  
(the middle panel) as well as the ratio between the total magnetic energy 
density and kinetic energy density (the bottom panel). 
}
\label{KineticEnergy}
\end{center}
\end{figure}

The kinetic energy densities in the steady state between the run {\sf StarTop} without 
magnetic field and the four runs with magnetic field are compared in Figure \ref{KineticEnergy}. 
The peak value of $E_k$ in {\sf StarB3} is increased by a factor of 3 compared with the value 
in {\sf StarTop}. Despite the large differences of the initial magnetic field strengths, the kinetic 
energy density in the turbulent state is only increased by $50\%$ from {\sf StarB1} 
to {\sf StarB3}. The ratios between the magnetic and kinetic energy densities in the turbulent state 
only vary from 0.2 to 0.6 in the three runs {\sf StarB1}, {\sf StarB2} and {\sf StarB3}, as shown in 
the bottom panel of Figure \ref{KineticEnergy}. 
This suggests that convection is driving a reasonably efficient small-scale magnetic dynamo. 
The run {\sf StarB4} reaches similar 
kinetic energy density and vertical component of magnetic energy density as in the run {\sf StarB3}, 
although the initial horizontal magnetic field is $5$ times stronger in {\sf StarB4} compared with 
{\sf StarB3}. This also suggests that the photon bubble instability that 
shows up in the run {\sf StarB4} (Figure \ref{snapshot_early}) 
does not affect the vertical structures of the envelope (such as density, kinetic 
energy density, entropy) significantly compared with the run {\sf StarB3}, although 
the two runs do have different ratios $E_B/E_k$.

\section{Radiation Force and Energy Transport}
\label{sec:result2}
\subsection{The Porosity Factor}
\label{sec:porosity}

\begin{figure}[htp]
\begin{center}
\includegraphics[width=1.0\columnwidth]{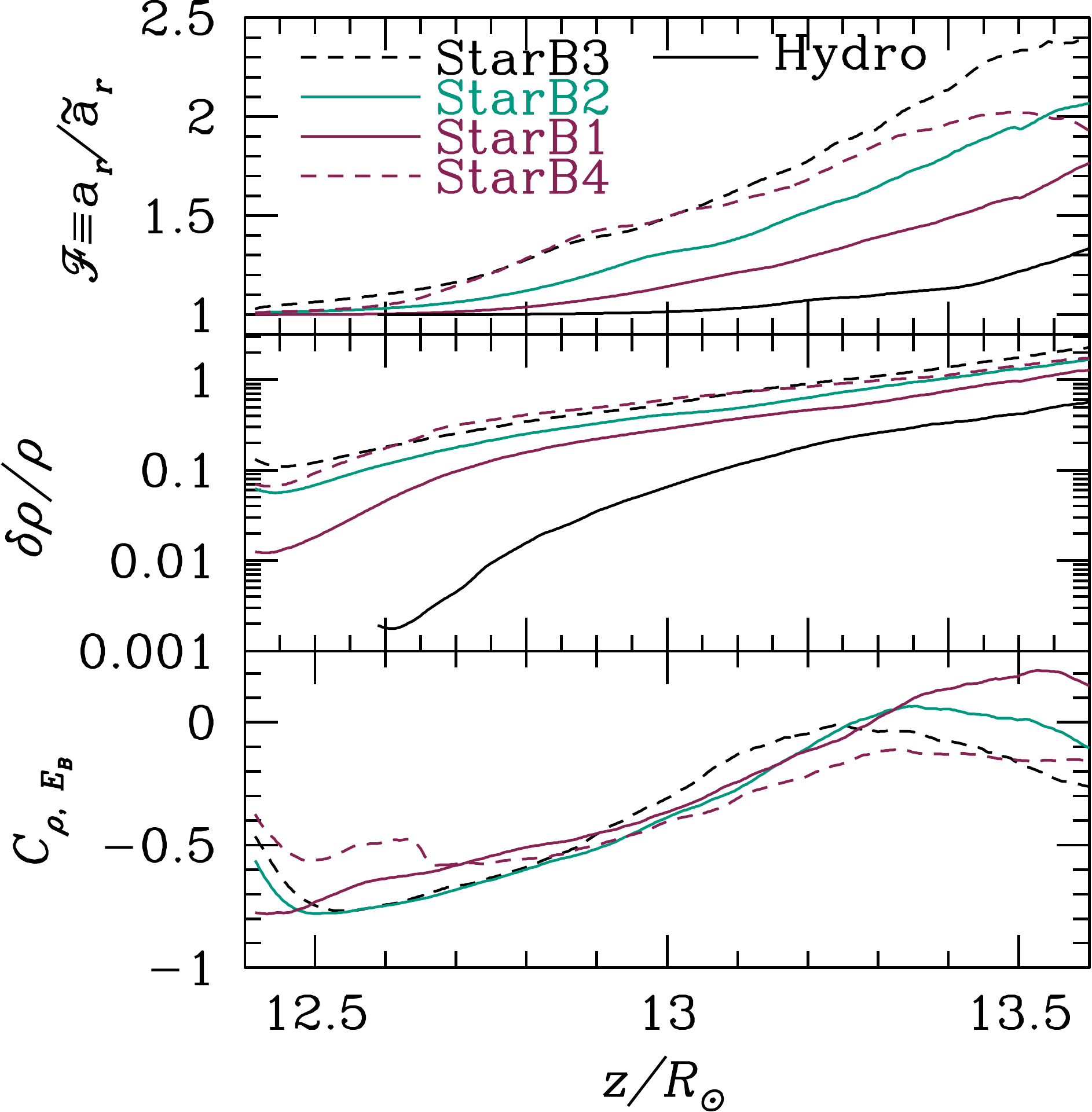}
\caption{Top: vertical profiles of the time averaged porosity factor, which is the ratio between the 
volume averaged and density weighted radiation acceleration $a_r$ and $\tilde{a}_r$. Middle: vertical profiles 
of the time averaged ratio between the standard deviation of density and mean density at each height. 
Bottom: time averaged vertical profiles of the cross correlations between density and magnetic energy 
density fluctuations.
Different lines represent the five simulations as indicated in the top panel of the figure. 
Porosity effects become larger with increasing magnetic fields because of larger density 
fluctuations. 
}
\label{porosity}
\end{center}
\end{figure}

The simulation {\sf StarTop} is in the regime that the optical depth per pressure scale height 
$\tau_0$ is smaller than the critical value $\tau_c$ \citep[][]{Jiangetal2015}. This is also true for all 
the MHD runs, even though the opacity peak is increased a little bit when the envelopes shrink. 
One important effect in this regime is that when photons go through the 
turbulent convective region in this regime, the radiation flux is stronger in the low density region \citep[e.g.,][]{Shaviv1998}. 
This causes an anti-correlation between density and radiation flux fluctuations at each height. The 
density weighted radiation acceleration $\tilde{a}_r$ (equation 14 of \citealt{Jiangetal2015}) can be 
decomposed as
\begin{eqnarray}
\tilde{a}_r&=&\frac{\langle(\rho-\langle\rho\rangle)\kappa_t F_{r,0z}\rangle+\langle\rho\rangle\langle\kappa_t F_{r,0z}\rangle}{c\langle\rho\rangle}=\frac{\langle\kappa_t F_{r,0z}\rangle}{c}\nonumber\\
&+&\frac{\left\langle(\rho-\langle\rho\rangle)(\kappa_t F_{r,0z}-\langle\kappa_t F_{r,0z}\rangle)\right\rangle}{c\langle\rho\rangle}.
\end{eqnarray}
Because of the existence of the anti-correlation, this is smaller than the volume averaged 
radiation acceleration $a_r=\langle\kappa_t F_{r,0z}\rangle/c$. The porosity factor 
\begin{eqnarray}
\text{\cF}\ \equiv\frac{a_r}{\tilde{a}_r}
\end{eqnarray}
for the hydro run {\sf StarTop} and the four MHD runs are shown in the top panel of Figure \ref{porosity}. 
Although magnetic pressure is small compared with the radiation pressure, it increases the porosity 
factor significantly compared with the hydro case. The stronger the initial magnetic field strength, 
the larger is the porosity factor. In the run {\sf StarB3}, {\cF} \ reaches $1.4$ at the iron opacity peak. 
It is also not sensitive to the initial horizontal magnetic field, as the two runs {\sf StarB3} and {\sf StarB4} 
have very similar  {\cF}.\ \ This is also consistent with the fact that the two runs have very similar 
vertical structures. The increased porosity factor with magnetic field explains why the envelopes with 
magnetic field will shrink and $\tilde{a}_r$ in the MHD runs is actually smaller than the value in the 
hydro case, although the volume averaged radiation accelerations in the MHD runs are larger. 

Magnetic field increases the porosity factor as it increases the density fluctuations, 
which is shown in the middle panel of Figure \ref{porosity}. 
The increase of $\delta\rho/\rho$ from {\sf StarTop} to {\sf StarB3} 
agrees pretty well with the change of {\cF}. At a given height, the low and high density 
regions should be roughly in pressure balance. In the hydro case, the gas pressure difference caused by the 
density fluctuations can only be balanced by the radiation pressure horizontally. But because of 
rapid diffusion, fluctuations of radiation pressure are small (Figure 20 of \citealt{Jiangetal2015}), 
which limits the amount of density fluctuations that can be achieved. However, with magnetic field, 
magnetic pressure is larger in the low density regions, which can balance the gas pressure 
gradient. This is confirmed in the bottom panel of Figure \ref{porosity}, which 
shows strong anti-correlations between density and magnetic energy density 
fluctuations (negative $C_{\rho, E_B}$) for all the MHD runs. Here $C_{\rho, E_B}$ is 
the cross correlation coefficient between $\rho$ and $E_B$ normalized by their standard 
deviations at each height. 
This allows larger density fluctuations compared with the hydro case. 
Therefore, the anti-correlation 
between density and radiation flux as well as the porosity factor {\cF}\  \ are increased.

\subsection{Vertical Energy Transport}

\begin{figure}[htp]
\begin{center}
\includegraphics[width=1.0\columnwidth]{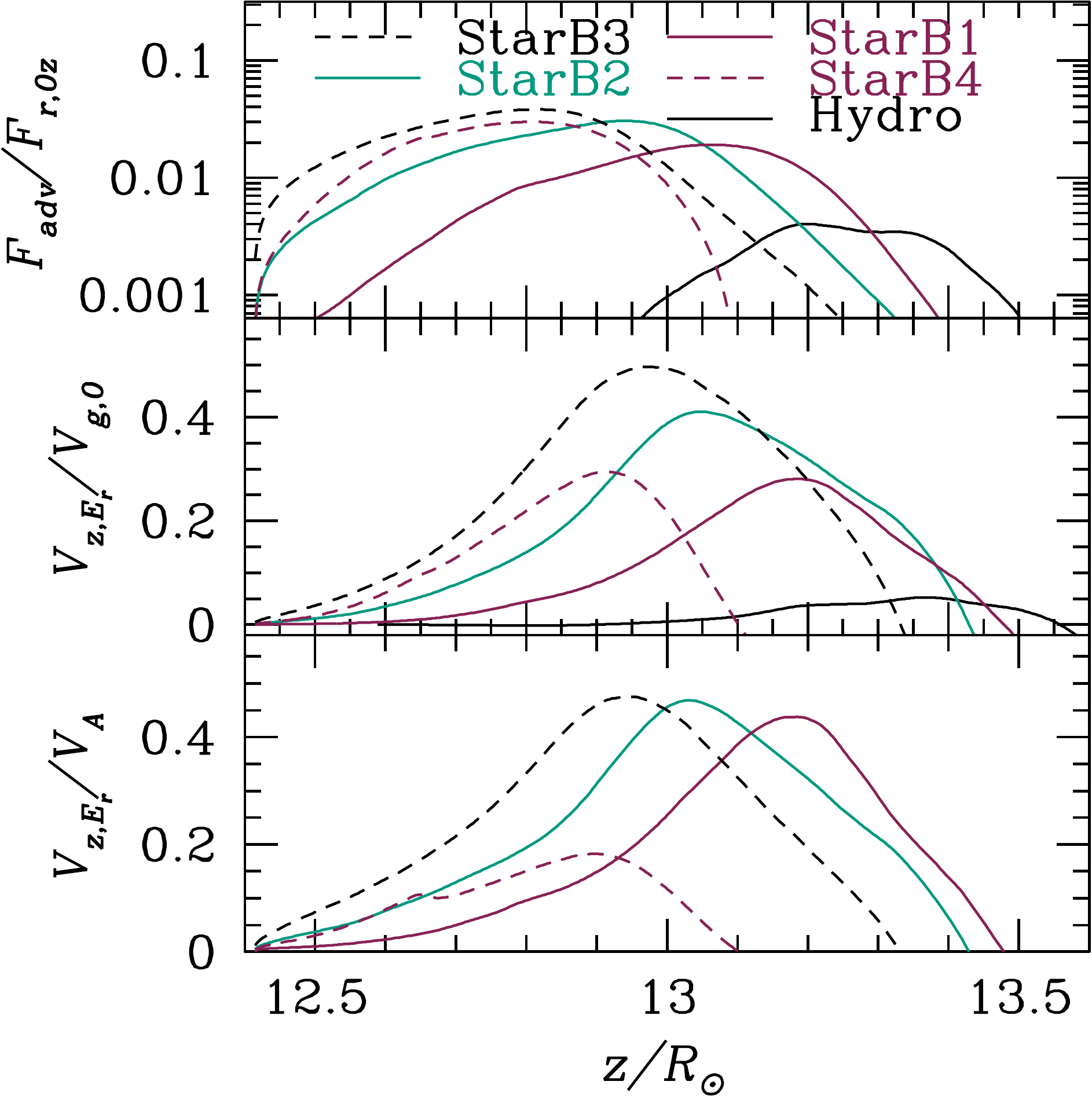}
\caption{Top: vertical profiles of the time and horizontally averaged advective energy flux $F_{adv}$
scaled by the total vertical radiation flux $F_{r,0z}$. Middle: vertical profiles of 
the time and energy weighted vertical advection velocity in unit of the isothermal 
sound speed at the fiducial temperature $T_0$. Bottom: vertical profiles of the ratio 
between the time averaged energy advection velocity and the time averaged Alfv\'en velocity. 
For a given magnetic field configuration, the ratios $V_{z,E_r}/V_A$ are very similar 
for a wide range of magnetic field strength. 
}
\label{energyflux}
\end{center}
\end{figure}

In the regime when $\tau_0<\tau_c$, most of the energy is transported by the diffusive radiation flux 
while the advective radiation flux $F_{\text{adv}}=v_zE_r$ is negligible. This is because the photon 
diffusion time scale across the pressure scale height is smaller than the sound crossing time for 
the isothermal sound speed in this regime. However, with magnetic fields, magnetic buoyancy increases 
the vertical advection velocity and thus the advection flux, which is shown in the top panel of Figure \ref{energyflux}. 
The ratio between the vertical advection flux and the total flux increases from $\sim0.4\%$ in {\sf StarTop} 
to $4\%$ in {\sf StarB3}. 
Although the advection flux is increased with magnetic field, 
it is still a small fraction ($<4\%$) of the total energy flux. Different vertical structures of the 
envelopes with magnetic field are not due to the change of advection flux. Instead, it is due to 
different porosity factors. 

The larger the initial vertical magnetic field is, the stronger the advection flux. 
The vertical advection velocity $V_{z,E_r}\equiv \langle V_zE_r\rangle/\langle E_r\rangle$ also increases with increasing 
magnetic field for a given magnetic field configuration as shown in the middle panel of Figure \ref{energyflux}. 
The effect of enhanced advection flux due to magnetic buoyancy is also observed in black hole accretion disk 
simulations with turbulence caused by magneto-rotational instability (MRI) \citep[][]{Blaesetal2011,Jiangetal2014}. 
As shown in Figure 9 of \cite{Blaesetal2011},  this can also be easily understood in terms of the larger density 
fluctuations due to magnetic field. The radiation energy density $E_r$, density $\rho$ and vertical 
velocity $v_z$ at each 
height can be separated into the time and horizontally averaged mean values and the fluctuation components
\begin{eqnarray}
E_r&=&\langle E_r\rangle+\delta E_r,\nonumber\\
\rho&=&\langle\rho\rangle+\delta \rho,\nonumber\\
v_z&=&\langle v_z\rangle+\delta v_z.
\end{eqnarray} 
Because there is no net mass flux, 
\begin{eqnarray}
\langle\rho v_z\rangle=\langle\rho\rangle \langle v_z\rangle+\langle\delta\rho\delta v_z\rangle=0.
\end{eqnarray} 
Therefore, we have
\begin{eqnarray}
\langle v_z\rangle=\frac{-\langle\delta\rho\delta v_z\rangle}{\langle\rho\rangle}.
\end{eqnarray} 
The vertical energy advection velocity can then be calculated as
\begin{eqnarray}
V_{z,E_r}&\equiv&\frac{\langle V_zE_r\rangle}{\langle E_r\rangle}=\langle v_z\rangle+\frac{\langle\delta E_r\delta v_z\rangle}{\langle E_r\rangle}\nonumber\\
&=&\left\langle\delta v_z\left(\frac{\delta E_r}{\langle E_r\rangle}-\frac{\delta \rho}{\langle\rho\rangle}\right)\right\rangle.
\end{eqnarray} 
As relative fluctuations of $E_r$ are much smaller than the relative fluctuations in $\rho$ in this rapid 
diffusion regime \citep[][]{Jiangetal2015}, the radiation advection flux 
is basically the mean vertical velocity $-\langle\delta \rho\delta v_z\rangle/\langle\rho\rangle$ multiplied by the 
horizontally averaged radiation energy density. That is why radiation advection flux increases 
with increasing density fluctuations. 
Comparing the two runs {\sf StarB3} and {\sf StarB4} shows that the vertical advection velocity does 
depend on the magnetic field configuration. Larger horizontal magnetic field actually decreases the 
vertical energy advection velocity.

Following \cite{StellaRosner1984} (see also \citealt{MacGregorCassinelli2003} and 
\citealt{CantielloBraithwaite2011}), we show that 
when the vertical energy advection velocity (and thus density fluctuation) 
is dominated by magnetic buoyancy, it is proportional to the Alfv\'en velocity $V_A$. 
Considering density fluctuations with typical radius $R$, magnetic pressure 
inside the low density region balances the gas pressure inside the high 
density region with almost a constant temperature at each height due to rapid 
photon diffusion. The buoyancy force per unit length is $\approx (\pi R^2/\lambda_p)P_B$, 
where $\lambda_p=H_0c_g^2/c_{r,s}^2$ is the gas pressure scale height. In steady state, 
the buoyancy force is balanced by the drag $\rho v_z^2RC_D$, where $C_D$ is the 
drag coefficient. Then the vertical velocity can be estimated as
\begin{eqnarray}
v_z^2\approx \frac{\pi R}{2\lambda_pC_D}V_A^2.
\end{eqnarray}
The ratio between the vertical energy advection velocity and Alfv\'en velocity for the MHD runs 
are shown in the bottom panel of Figure \ref{energyflux}. For runs {\sf StarB1}, {\sf StarB2} and 
{\sf StarB3}, which have the same initial magnetic field configuration, the ratios
$V_{z,E_r}/V_A$ are pretty similar despite the large differences of magnetic field strength. 
The ratios peak at the location of the iron opacity peaks for each run. 
The coefficient $R/(\lambda_pC_D)$ decreases with distance from the iron opacity peak.  
It also depends on the magnetic field configuration as shown by the run {\sf StarB3} and {\sf StarB4}. 
Its value is smaller with stronger horizontal magnetic field. 




\subsection{Magnetic Fields in the Efficient Convection Regime}
In the regime $\tau_0\gg\tau_c$ as in the run {\sf StarDeep} studied 
by \cite{Jiangetal2015}, the porosity factor \cF \ \ is $1$ (see panel d of Figure 7 
in \citealt{Jiangetal2015}), which means that the effective radiation acceleration is 
not reduced by the density and radiation flux fluctuations. With magnetic fields, the density 
fluctuations are expected to be larger, but as long as $\tau_0$ in the low density region is 
still much larger than $\tau_c$, \cF \ \ is not going to be affected.

In the regime of efficient convection ($\tau_0\gg\tau_c$),  magnetic buoyancy will still increase the vertical 
advection velocity, but it cannot increase the advection flux significantly, 
as this is limited by the total available energy flux. However, 
magnetic buoyancy may result in different 
vertical structures compared to the case of efficient hydrodynamic convection. 
For example, we have shown that the advection flux caused by magnetic 
buoyancy is proportional to $V_AE_r$, which cannot be calculated 
based on the thermal entropy gradient as in the mixing length theory. 
Detailed comparisons between the magnetic buoyancy driven 
convection and mixing length theory, as well as the resulting 
stellar structures, will be the subjects of future investigation.

\section{Discussions and Conclusions}
\label{sec:discussion}
We have extended the work of \cite{Jiangetal2015} by studying the effects of magnetic fields on the envelope structure and energy
transport of  main sequence massive stars at the location of the iron opacity peak. 
We focus on the regime $\tau_0\ll \tau_c$ so that the diffusion time scale is shorter than the local dynamical
time scale and convection is inefficient. 
We found that the presence of magnetic fields with amplitude $\sim 100$~G-1~kG increases the density fluctuations  
in the turbulent envelope, resulting in a larger porosity 
factor and consequentially an enhanced transport of energy by the radiation field.
The presence of magnetic buoyancy  also increases the advection flux 
significantly, with the energy advection velocity proportional to the Alfv\'en 
velocity for a fixed magnetic field configuration. 
Both effects contribute to increasing the energy transport across the turbulent envelope compared 
to the pure hydrodynamic case, which causes the envelope to shrink by several scale heights.

\subsection{Implications for 1D Stellar Evolution}
Stellar evolution calculations adopt mixing length theory for calculating the energy transport in inefficient convective regions. 
In this regime, different assumptions for the mixing length $\alpha H$ result in large differences in the stellar radii of massive star models 
\citep[see e.g. Fig.~20 in][]{Koheler2015}. However, as discussed in e.g., \citet{Jiangetal2015}, mixing length theory is inadequate for treating 
the energy transport in radiation dominated stellar envelopes when convection is inefficient ($\tau_0<\tau_c$), meaning that current stellar 
evolution calculations may provide inaccurate results.  
This is also true in the presence of magnetic fields. 
Because the entropy 
gradient is smaller in the MHD runs compared with the hydro run {\sf StarTop} (Figure \ref{Profile}), 
if we adopt the same $\alpha=0.4$ as in {\sf StarTop}, the convective fluxes predicted by the mixing length theory 
for {\sf StarB1}, {\sf StarB2} and {\sf StarB3} are much smaller than the corresponding value for {\sf StarTop}. 
They are also significantly smaller than the advective flux we get from the simulations. In order to match 
the calculated fluxes with the mixing length theory at the iron opacity peaks, we need $\alpha=0.6, 0.8, 1$ 
for {\sf StarB1}, {\sf StarB2} and {\sf StarB3} respectively, which are much larger than the ratios between the 
correlation lengths and pressure scale height. Even if we can match the convective flux at the iron opacity peak, 
the convective flux drops quickly away from the iron opacity peaks in the mixing length theory 
and becomes much smaller than what we get from the simulations.  
To model inefficient convection  correctly in 1D stellar evolution codes, 
two steps are required: 1) Calibrating the mixing length theory 
using 3D MHD calculations, and 2) Including the effect of porosity on the radiative energy transport.

\subsection{Observational Consequences}
Inefficient convection causes the development of supersonic turbulent velocities with respect to the isothermal sound speed.  
As a result, we find large density and velocity fluctuations  in the simulated stellar envelopes.
The vertical velocity fluctuations increase with the initial amplitude of the 
vertical component of the magnetic field. The average turbulent velocity reaches $\sim 100\kms $
at the photosphere in our model with $B_{z,0}\sim 400$~G ({\sf StarB3}), which is about 
$7\%$ of the escape velocity. These fluctuations can impact the spectroscopic measurements 
of line profiles, potentially affecting both the microturbulence and macroturbulence parameters \citep{Cantiello2009,Jiangetal2015}.  
Recent spectroscopic observations of a large sample of OB stars revealed a very interesting trend, with macroturbulent 
velocities increasing with stellar luminosity \citep{SimonDiaz:2016}. This has been tentatively attributed to the increasing 
turbulent pressure in the iron convection zone \citep{Grassitelli2015,Grassitelli:2016}, since turbulent pressure fluctuations 
may trigger high-order high-angular degree oscillations that can collectively mimic the effect of macroturbulence \citep{Aerts2009}. 
Despite capturing the relevant physics, our cartesian box calculations can not provide detailed predictions for the observed 
spectroscopic signature of radiation dominated envelopes of massive stars. Such predictions require global envelope calculations, 
which we will explore in future work. 

Here we have studied the effects of magnetic fields up to values close to equipartition with thermal pressure at the location of the 
iron opacity peak ($\sim$ kG). Magnetic fields with  larger amplitude are expected to have a  different impact on the convective properties, 
potentially decreasing the amount of turbulence in the outer stellar envelope. Such effects were discussed by \citet{Sundqvist:2013},
who reported the observation of magnetic inhibition of photospheric macroturbulence in an O-star with a  surface dipolar field of
 $\sim$ 20 kG. Interestingly, magnetic O-stars with smaller amplitude surface magnetic fields (implying sub-equipartition with thermal
  pressure at the location of the iron opacity peak) show normal macroturbulent velocities (20-60 $\kms$).  This result strongly 
  supports a causal link between the observed surface turbulence and the sub-surface convection associated with the iron
  opacity peak \citep{Cantiello2009,Sundqvist:2013,Grassitelli2015,Grassitelli:2016}.

It has been suggested that  magnetic fields can be generated by dynamo action in the sub-surface convective regions \citep{Cantiello2009,Cantiello:2011}, 
and that they can reach the photosphere and affect the observable properties of massive stars 
\citep{CantielloBraithwaite2011}. Our calculations support these claims, showing that an initial magnetic field is efficiently 
amplified by turbulent motions around the iron opacity peak and rises buoyantly to reach the stellar photosphere (see Fig.~\ref{energy}). 
Possible impact of these buoyant magnetic fields includes photometric variability due to surface magnetic spots and/or 
prominences \citep{CantielloBraithwaite2011,Sudnik:2016}, excess X-ray emission 
\citep{Gagneetal1997,BabelMontmerle1997,Waldron:2009,Nazeetal2014,Owockietal2016}, 
as well as wind variability and clumping \citep{Puls:2006,Michaux:2014}. 

While massive stars are rapid rotators, in this work we have not considered the effects of rotation.
It is conceivable that the inclusion of rotation might impact the details of dynamo action
in the turbulent convection around the iron opacity peak. Moreover, even in the radiative parts 
of the stellar envelope, differential rotation is expected to potentially trigger the Spruit-Tayler dynamo 
\citep[][]{Spruit2002} or even the MRI \citep[e.g.][]{Wheeleretal2015}. The interplay between these different 
dynamo-generated magnetic fields could be extremely important for angular momentum transport 
and internal mixing across massive star envelopes. These local calculations also do not include the possibility 
of mass loss, which may also change the structures in the iron opacity region \citep[][]{RoMatzner2016}. 
The interplay between convection in this region and winds will be the focus of future global calculations of massive star 
envelopes.

 

\section*{Acknowledgements}
Y.F.J thanks all members
of the SPIDER network for helpful discussions. This work was supported
by the computational resources provided by the NASA High-End Computing
(HEC) Program through the NASA Advanced Supercomputing (NAS) Division
at Ames Research Center; and the National Energy Research Scientific
Computing Center, a DOE Office of Science User Facility supported by
the Office of Science of the U.S.  Department of Energy under Contract
No. DE-AC02-05CH11231.  
This research is funded in part by the Gordon and Betty Moore Foundation through Grant 
GBMF5076 to L. B. and E. Q. and is supported in part by the National Science Foundation 
under NSF PHY-1125915 and  by NASA under NNX14AB53G. 
EQ was also supported in part by a Simons Investigator Award from the Simons Foundation.

\bibliography{MassiveStarMHD}

\end{CJK*}

\end{document}